\documentclass[aip,apl,reprint,noshowpacs,noshowkeys,amssymb,amsmath,amsfonts,bibnotes]{revtex4-2}

\usepackage{graphicx}
\usepackage{theorem}
\usepackage{dcolumn}
\usepackage{longtable}
\usepackage{bm}
\usepackage{accents}
\usepackage{color}
\usepackage[dvipsnames]{xcolor}
\usepackage[normalem]{ulem}
\usepackage{multirow}

\renewcommand{~}{\,}

\makeatletter
\renewcommand{\figurename}{Fig.}
\renewcommand{\tablename}{Tab.}
\renewcommand{\fnum@figure}[1]{\textbf{\figurename~\thefigure.} }
\renewcommand{\fnum@table}[1]{\textbf{\tablename~\thetable.} }
\makeatother

\begin{document}

\title{Solid neon as a noise-resilient host for electron qubits above 100 mK}

\author{Xinhao Li}
\affiliation{Center for Nanoscale Materials, Argonne National Laboratory, Lemont, Illinois 60439, USA\looseness=-1}
\affiliation{Department of Physics, Harvard University, Cambridge, Massachusetts 02138, USA\looseness=-1}

\author{Christopher S. Wang}
\affiliation{James Franck Institute and Department of Physics, University of Chicago, Chicago, Illinois 60637, USA\looseness=-1}

\author{Brennan Dizdar}
\affiliation{James Franck Institute and Department of Physics, University of Chicago, Chicago, Illinois 60637, USA\looseness=-1}

\author{Yizhong Huang}
\affiliation{Center for Nanoscale Materials, Argonne National Laboratory, Lemont, Illinois 60439, USA\looseness=-1}
\affiliation{Department of Electrical and Computer Engineering, Northeastern University, Boston, Massachusetts 02115, USA\looseness=-1}

\author{Yutian Wen}
\affiliation{Department of Physics and Astronomy, University of Notre Dame, Notre Dame, Indiana 46556, USA\looseness=-1}

\author{Wei Guo}
\affiliation{National High Magnetic Field Laboratory, Tallahassee, Florida 32310, USA}
\affiliation{Department of Mechanical Engineering, FAMU-FSU College of Engineering, Florida State University, Tallahassee, Florida 32310, USA}

\author{Xufeng Zhang}
\affiliation{Department of Electrical and Computer Engineering, Northeastern University, Boston, Massachusetts 02115, USA\looseness=-1}

\author{Xu Han}\email[Email: ]{xu.han@anl.gov}
\affiliation{Center for Nanoscale Materials, Argonne National Laboratory, Lemont, Illinois 60439, USA\looseness=-1}
\affiliation{Pritzker School of Molecular Engineering, University of Chicago, Chicago, Illinois 60637, USA\looseness=-1}

\author{Xianjing Zhou}\email[Email: ]{xianjing.zhou@fsu.edu}
\affiliation{National High Magnetic Field Laboratory, Tallahassee, Florida 32310, USA}
\affiliation{Department of Mechanical Engineering, FAMU-FSU College of Engineering, Florida State University, Tallahassee, Florida 32310, USA}

\author{Dafei Jin}\email[Email: ]{dfjin@nd.edu}
\affiliation{Center for Nanoscale Materials, Argonne National Laboratory, Lemont, Illinois 60439, USA\looseness=-1}
\affiliation{Department of Physics and Astronomy, University of Notre Dame, Notre Dame, Indiana 46556, USA\looseness=-1}

\date{\today}

\begin{abstract}

\end{abstract}

\maketitle
\pretolerance=9000 

\textbf{Solid neon can be used as a solid host for single-electron qubits, and at temperatures of around 10~mK, electron-on-solid-neon charge qubits exhibit long coherence times and high operation fidelities. However, systematic characterization of the noise features of such systems is needed for the development of scalable quantum information architectures. Here, we show that solid neon can be used as a noise-resilient host for electron qubits above 100~mK. We examine the resilience of solid neon against charge and thermal noise when electron-on-solid-neon charge qubits are operated away from the charge-insensitive sweet spot and at elevated temperatures. We show that the extracted high-frequency charge noise density of electron-on-solid-neon qubits, projected as voltage fluctuations on nearby electrodes, is between $10^{-4}$ and $10^{-6}~\mathrm{\mu V^2/Hz}$ at 0.01 to 1 MHz, which is comparable with common semiconductor hosts. We also show that the electron-on-solid-neon charge qubits operating around 5 GHz frequencies can maintain echo coherence times of over 1 $\mu$s at temperatures up to 400 mK.}

Solid-state electron qubits are subject to various decoherence channels from their host materials. Noise spectral analysis shows a characteristic $1/f$ behaviour for electron charge or spin qubits, typically attributed to charge fluctuators in the bulk or at the interfaces of the host materials~\cite{paladino20141,burkard2023semiconductor}.
The coherence of electron qubits can be extended by reducing the noise density in their hosting environment and minimizing their susceptibility to noise~\cite{de2021materials, burkard2023semiconductor}.
We recently showed that solid neon can be a host material to trap electrons at a vacuum/neon interface~\cite{zhou2022single, zhou2024electron}.
The charge states of electrons on solid neon can be addressed by coupling them with superconducting resonators in a circuit quantum electrodynamics (cQED) architecture.
At the charge sweet spot, they are first-order insensitive to charge noise, exhibiting long coherence times $T_2^*$ up to $\sim$50\,\textmu s (ref.~\cite{zhou2024electron}). 
This is nearly four orders of magnitude longer than that of semiconductor charge qubits~\cite{petersson2010quantum}.
The long coherence of electron-on-solid-neon (eNe) charge qubits also leads to long coherence eNe spin qubits~\cite{chen2022electron, guo2024quantum, jennings2024quantum}.

However, eNe charge qubits are susceptible to environmental decoherence channels.
Thus, understanding the performance of eNe qubits away from the sweet spot and at elevated temperatures, where they are subject to charge and thermal noise, will be important in improving qubit uniformity, crucial for the generation and retention of entangled states~\cite{paladino20141}.
Such studies are also required for the development of eNe spin qubits~\cite{jennings2024quantum, chen2022electron, tian2025nbtin} through electrically sensitive mechanisms, such as exchange interactions~\cite{eng2015isotopically,jock2022silicon, connors2022charge, cerfontaine2016feedback,  dial2013charge, cerfontaine2020closed} and synthetic spin–orbit coupling~\cite{klemt2023electrical,zwerver2022qubits, yoneda2018quantum, struck2020low, kawakami2016gate}.
Furthermore, operating qubits at elevated temperatures can mitigate engineering constraints created due to the low cooling power at millikelvin temperatures and can aid the scaling up of the technology~\cite{anferov2024superconducting, huang2024high}.

\begin{figure}[ht]
\centerline{\includegraphics[scale=0.16]{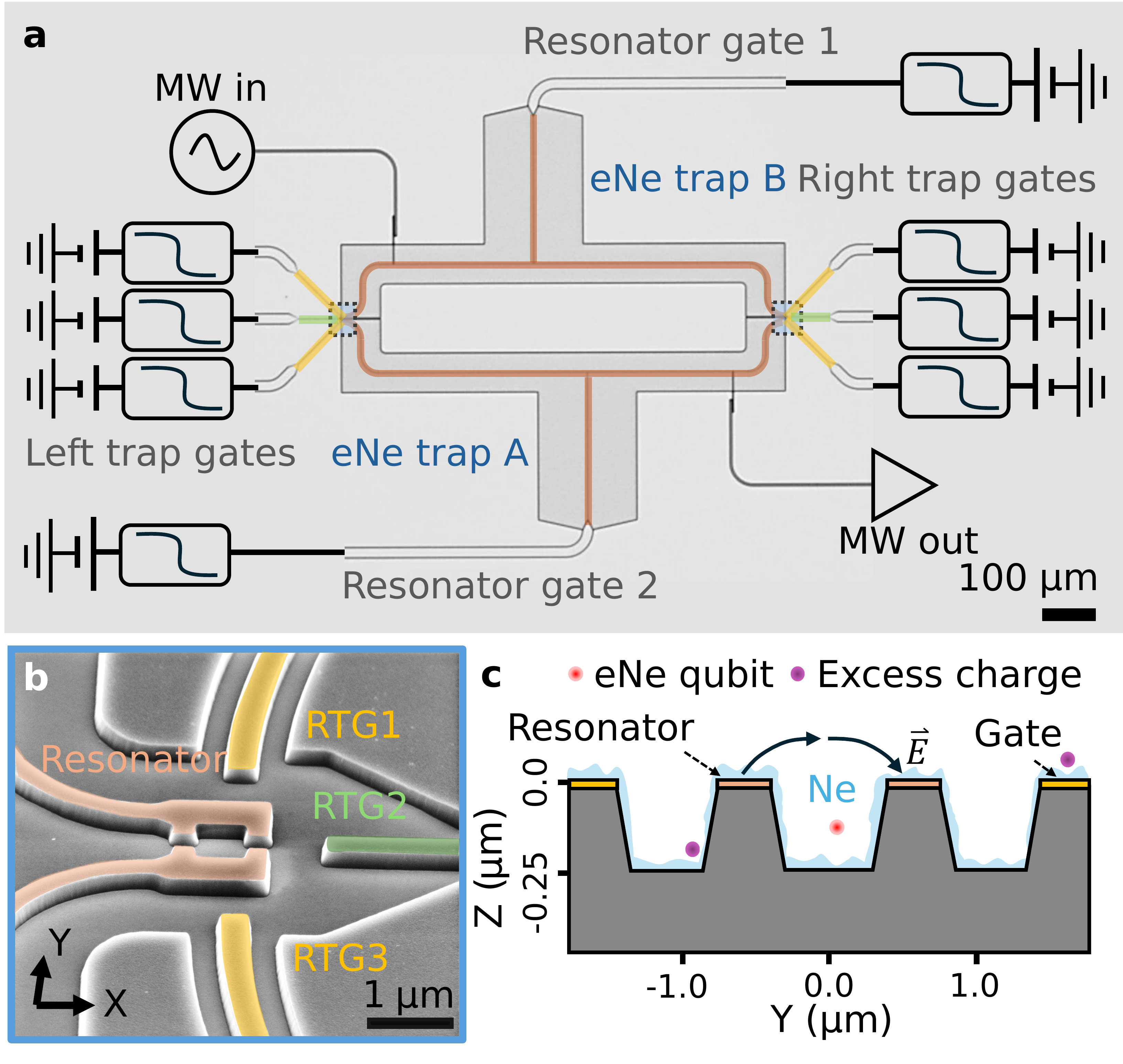}}
\caption{\textbf{eNe charge qubit coupled to a TiN high-impedance superconducting resonator}. \textbf{a}, Illustration of the high impedance TiN superconducting resonator with two identical electron traps, microwave (MW) input and output couplers, and direct current (DC) gates. The metal plane between the resonator pins is connected to the ground plane via aluminium wire bonds. \textbf{b}, False-colour scanning electron micrograph image of the electron trapping area on the right side of the resonator with trap gates (RTG). Resonator bias voltage $V_{\text{res}}$ is applied symmetrically on the two pins of the resonator via the resonator gates in a. \textbf{c}, Cross-section schematic of the electron trapping area following the white dashed line in b, where the charge state of eNe qubit is coupled with the MW electric field $\vec{E}$ within the resonator.} \label{Fig:Device}
\end{figure}

In this Article, we report the characterization of the noise features of solid neon.
We use eNe qubits to probe the environmental noise on solid neon films, exploiting the increased charge sensitivity of the qubits when biased away from the sweet spots.
We show that the extracted high-frequency charge noise density of eNe, projected as voltage fluctuations on nearby electrodes, is between $10^{-4}$ and $10^{-6}~\mathrm{\mu V^2/Hz}$ at 0.01 to 1 MHz.
This is lower than some common semiconductor hosts
~\cite{jock2022silicon, zwerver2022qubits, klemt2023electrical, connors2022charge, yoneda2018quantum, struck2020low, kawakami2016gate,eng2015isotopically,cerfontaine2016feedback, hendrickx2021four,stehouwer2024exploiting, hendrickx2024sweet}, and approaches the best noise records achieved on GaAs/AlGaAs platforms ($10^{-6}$ to $10^{-7}~\mathrm{\mu V^2/Hz}$; ref~\cite{dial2013charge,cerfontaine2020closed}). 
We also show variations in noise density and qubit properties, which could be attributed to disorders on solid neon films that trap excess electrons and open additional decoherence channels.
Furthermore, we show that eNe qubits operated near  $\sim$5\,GHz can maintain echo coherence times exceeding 1 µs at temperatures of up to 400\,mK, and are primarily limited by thermally increased energy relaxation and dephasing rates.\\

\begin{figure*}[htb]
\centerline{\includegraphics[scale=0.31]{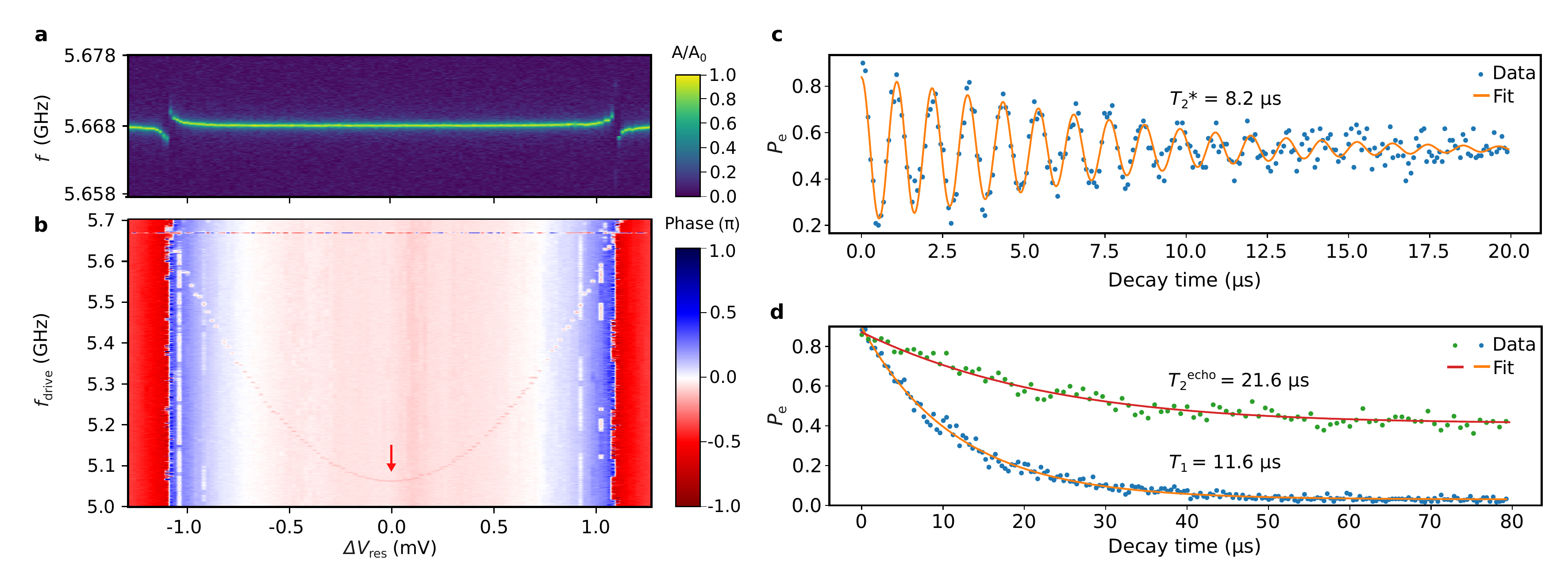}}
\caption{\textbf{Spectroscopic and coherence properties of an eNe charge qubit Q1}. \textbf{a}, Normalized microwave transmission amplitude ($A/A_0$) centred around the resonator frequency versus the relative resonator bias voltage $\Delta V_{\text{res}}$ as described in b. Two avoided crossings appear when the eNe qubit comes into resonance with the resonator. \textbf{b}, Two-tone qubit spectroscopy measurement displaying the transmission phase response at the resonator frequency $\omega_\text{r}$ versus $\Delta V_{\text{res}}$, while weak drive tones concurrently sent in at a frequency of $f_{\text{drive}}$. \textbf{c}, Ramsey fringes at the charge sweet spot, marked with the red arrow in b, with fitted total dephasing (Ramsey) time $T_2^*$ of 8.2\,\textmu s. $P_\text{e}$ is the qubit's excited-state population. \textbf{d}, Relaxation and Hahn echo measurements showing total decoherence time with a Hahn echo $T_2^{\text{echo}}$ = 21.6\,\textmu s and relaxation time $T_1$ = 11.6\,\textmu s at the charge sweet spot.}
\label{Fig:Q1_performance}
\end{figure*}

\noindent\textbf{Device structure}

Our device consists of a split superconducting resonator made of a 30-nm-thick titanium nitride (TiN) film grown on a ⟨111⟩-oriented intrinsic silicon substrate by atomic layer deposition~\cite{shearrow2018atomic} (Fig.~\ref{Fig:Device}a).
An electron trap is at each end of the resonator, surrounded by direct current (DC) gates to tune the qubit frequency, as shown in Fig.~\ref{Fig:Device}b. 
The differential mode of the resonator is coupled with the motional states of electrons trapped on solid neon, with the microwave electrical field pointing from one resonator pin to the other, as shown in Fig.~\ref{Fig:Device}c.
Considering the first two charge states of eNe, the coupled system can be described by the Jaynes-Cummings Hamiltonian~\cite{schuster2010proposal}:
\begin{equation}
 H=\hbar\omega_{\text{r}}\left(a^{\dagger}a+\frac{1}{2}\right)+ \frac{1}{2} \hbar \omega_{\text{q}} \sigma_z+g(a^{\dag}\sigma_{-}+a\sigma_{+}),
\end{equation}
where $\omega_{\text{r}}/2\pi$ = 5.668\,GHz is the resonator frequency after neon deposition, $\omega_{\text{q}}/2\pi$ is the qubit transition frequency, $g$ is the electron-photon coupling strength, $a^{\dag}$ and $a$ are the photon creation and annihilation operators, respectively, $\sigma_z$ and $\sigma_\pm\equiv \sigma_x\pm i\sigma_y$ are the standard Pauli operators on a two-level system.

As an improvement from previous work, we utilize the high kinetic inductance ($\sim$20~pH/$\Box$) of the thin TiN film to enhance the qubit-resonator coupling strength~\cite{koolstra2025high,tian2025nbtin}. The trap with a smaller size compared to the previous device~\cite{zhou2022single,zhou2024electron} further enhances the microwave field strength in the electron trapping area.
The estimated equivalent lumped element impedance for the differential mode is $Z_{\text{r}}\sim600~\Omega$ (Supplementary Information Section 1), approximately ten times that of the previous niobium (Nb) device~\cite{zhou2022single}.
Since $g\propto\omega_\text{r}\sqrt{Z_\text{r}}$, we expect the high impedance resonator to reach a coupling strength $g/2\pi\sim10~\text{MHz}$ level~\cite{koolstra2025high}. 
For high-impedance resonators, the parasitic capacitance of DC gates surrounding the trapping area can be comparable to the resonator capacitance~\cite{harvey2020chip}, causing microwave leakages. 
To minimize that, all gates are equipped with on-chip low-pass filters with a $\sim$0.5~GHz cut-off frequency providing over 60~dB attenuation at the resonator frequency.
With this design, the resonator maintains a narrow linewidth $\kappa/2\pi$ of 0.38\,MHz (Supplementary Information Section 1).

A thin layer of neon is grown on the device surface, giving less than 1\,MHz redshift of the resonator frequency, on which electrons generated from a tungsten filament are bound to. (See Methods for details.) 
At the trap, eNe qubits that strongly couple with the resonator exhibit vacuum Rabi splitting when the qubit frequencies are tuned across the resonator with the DC electrode. 
As shown in Fig.~\ref{Fig:Q1_performance}a, for one of the three qubits we characterized, labelled as Q1, the coupling strength extracted from the splitting gives $g/2\pi = 6.43$~MHz. 
Compared to the previous Nb resonator~\cite{zhou2022single,zhou2024electron}, the higher impedance of the TiN resonator enhances the qubit-resonator coupling strength, with a maximum observed $g/2\pi$ of approximately $16$~MHz. (See Supplementary Information Sections 2 to 5 for detailed characterization of the three qubits Q1, Q2, and Q3.)

\begin{figure*}[htb]
\centerline{\includegraphics[scale=0.3]{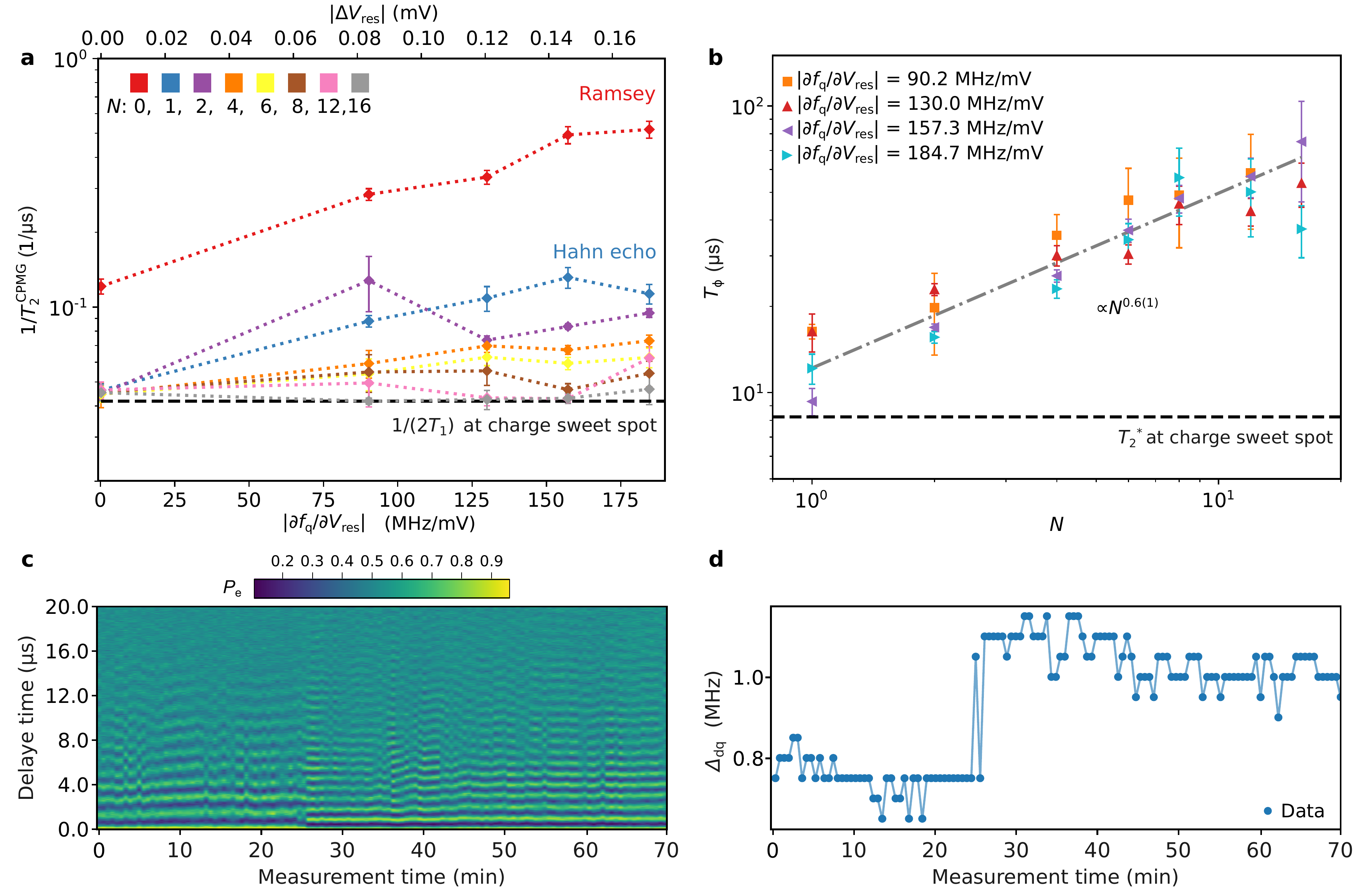}}	
\caption{\textbf{Decoherence of eNe qubit Q1}. \textbf{a}, Extracted total decoherence rate under CPMG sequences $1/T_{2}^{\rm{CPMG}}$ from experiments with Qubit 1 (Q1) biased at various frequency lever-arm $|\partial f_\mathrm{q}/\partial V_\mathrm{res}|$, and with refocusing pulse numbers $N =$ 0, 1, 2, 4, 6, 8, 12, and 16, denoted by different colours. With the increase of $N$, the $1/T_{2}^{\rm{CPMG}}$ of the qubit biased near the charge sweet spot approaches the limit of $2T_1$. \textbf{b}, The pure dephasing time $T_{\phi}$ increases as a function of $N$ when the qubit biased away from charge sweet spot, with a power-law fit of $T_{\phi} \propto N^{0.6(1)}$. \textbf{c}, Repeated Ramsey fringes measured near qubit sweet spot for 128 iterations, with each record taking 33~s. \textbf{d}, Detuning $\Delta_{dq}$ between drive tone and qubit frequency during the Ramsey measurements, revealing stochastic frequency shifts. All error bars represent the one standard error of extracted parameters (Methods).} \label{Fig:Dephasing}
\end{figure*}

\noindent\textbf{Noise sensitivity}

We utilize eNe qubits as sensitive probes to characterize the environmental charge noise.
Their sensitivity depends on the qubit spectral property, which can be mapped via two-tone measurements while varying the DC gate voltages, as shown in Fig.~\ref{Fig:Q1_performance}b for Q1.
The extracted qubit frequency follows a hyperbolic dependence on $V_{\text{res}}$ applied to the resonator with a charge sweet spot at 5.065~GHz (Supplementary Information Section 3).
We approximate the eNe qubit's transition frequency with a general model, capturing the measured qubit spectroscopic features:
\begin{equation}
\hbar\omega_{\text{q}}~=~\sqrt{\hbar^2(\omega_{\text{ss}}+\delta\omega_{\text{ss}})^2+(\epsilon+\delta\epsilon+2d_\text{e}(\mathcal{E}+\delta \mathcal{E}))^2},
\end{equation}
where $\omega_{\text{ss}}$ represents the charge sweet spot frequency, and $\epsilon$ describes the energy off-set defining the corresponding bias voltage.
The electrical tunability of the qubit's transition frequency is described by the term $2d_{\rm{e}}\mathcal{E}$, where $d_{\rm{e}}$ is the electron dipole moment and $\mathcal{E} \propto \Delta V_\text{res}$ is the applied field~\cite{hung2022probing, hegedus2025situ}.
Meanwhile, the noise terms $\delta \mathcal{E}$, $\delta \epsilon$ and $\delta\omega_{\text{ss}}$ caused by DC bias or adjacent charge fluctuations lead to qubit decoherence. The qubit's sensitivity to voltage fluctuations on the resonator electrodes can then be extracted as the frequency lever-arm $|\partial f_{q}/\partial V_\text{res}|$.

When the qubit is biased at the charge sweet spot, $|\partial f_{q}/\partial V_\text{res}|=0$, it is first-order insensitive to voltage (charge) noise.
Figure \ref{Fig:Q1_performance}c-d show the measured total dephasing (Ramsey) time $T_2^*$, and total decoherence time with a Hahn echo $T_2^{\text{echo}}$ of 8.2~\textmu s, and 22.6~\textmu s, respectively. 
The fact that $T_2^{\rm{echo}}$ approaches two times of the relaxation time $T_1$ indicates that the low-frequency (quasi-static) noise is the dominant dephasing factor for Q1 at its charge sweet spot~\cite{bylander2011noise}. 
To reveal the high-frequency (non-static) noise distribution, we need to bias the qubit away from its charge sweet spot.
The frequency lever-arms at the bias points used in the following noise analysis are $|\partial f_{q}/\partial V_\text{res}|=90.2$, $130.0$, $157.3$, and $184.7$ MHz/mV. 

Additionally, the sensitivity to gate voltage fluctuations varies between qubits (Supplementary Information Section 2).
As illustrated in Fig.~\ref{Fig:Device}c, the local neon profile and the trap structure define the potential energy landscape seen by the electrons.
Recent theoretical work has also emphasized the important role of neon morphology in defining the qubit's Hamiltonian~\cite{kanai2024single, zheng2025surface}.
The observed variation in qubit sensitivity to DC biases and their sweet spot frequencies confirms the non-uniformity in the local electron trapping potential.\\

\begin{figure*}[htb]
\centerline{\includegraphics[scale=0.3]{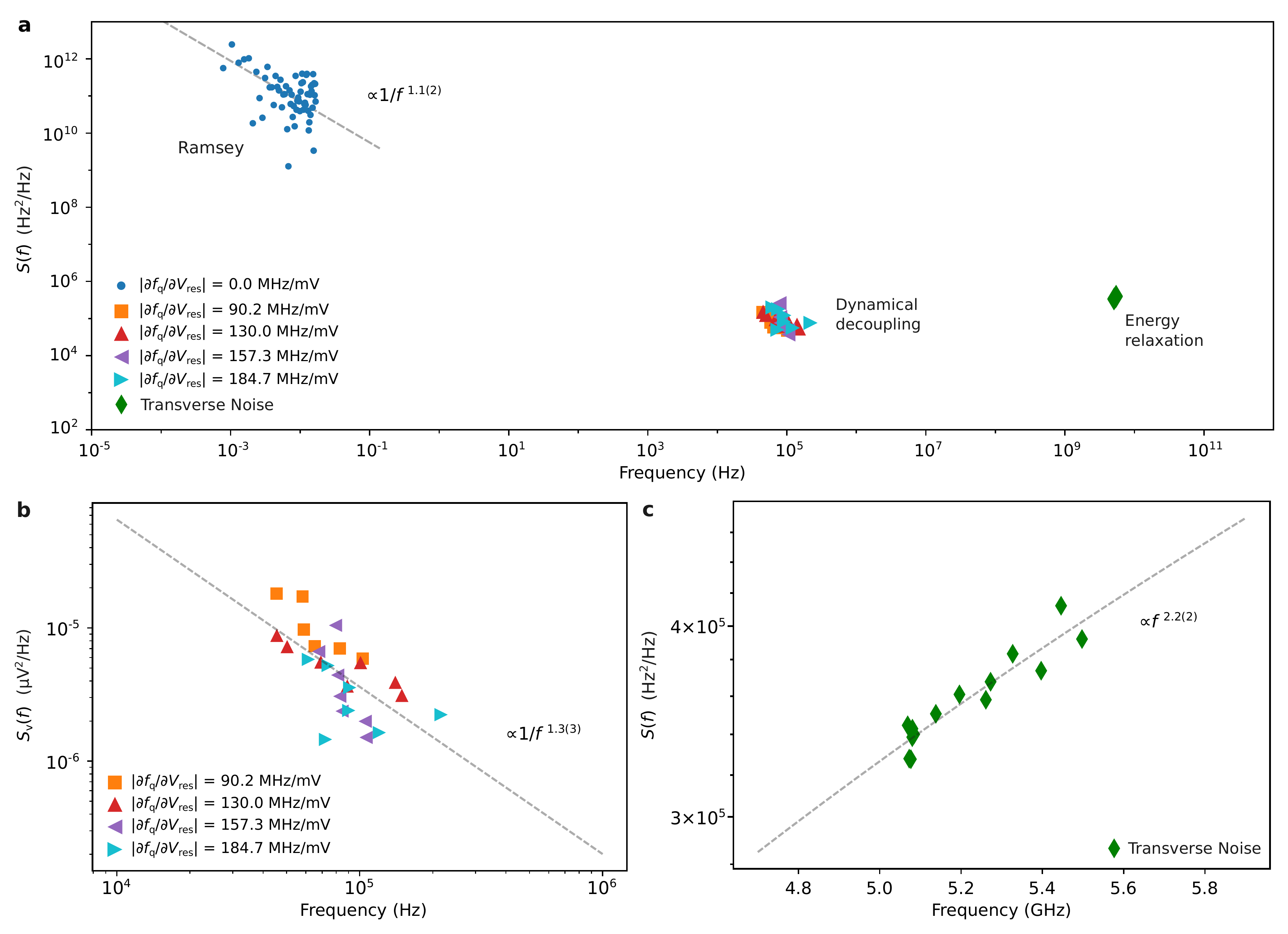}}
\caption{\textbf{Noise spectroscopy of eNe qubit Q1}. \textbf{a}, Data between 0.01 $\sim$ 1\,MHz: Total longitudinal noise density  (coloured dots) derived from dynamical decoupling data at different qubit bias points. Data between $10^{-3}$ to $10^{-1}$\,Hz: Extracted total longitudinal noise (blue dots) from long-term Ramsey measurements when biased at charge sweet spot. Data near 5.0\,GHz: Transverse noise of the eNe qubit (green diamonds). $S(f)$ represents the noise spectral density with $f$ denoting the frequency. \textbf{b}, Voltage noise density $S_\mathrm{v} (f)$ between 0.01 and 1\,MHz, projected on the resonator electrode. \textbf{c}, Zoom-in of the transverse noise in \textbf{a} between 5.0 and 5.6\,GHz. Notice that the x-axis is linear with the unit of GHz. Gray dashed lines: Power-law fits of frequency-dependent noise.} \label{Fig:Noise}
\end{figure*}

\noindent\textbf{Noise spectroscopy}

Fluctuations in the charge environment can cause stochastic frequency shifts of the eNe qubit, which lead to qubit dephasing. As illustrated in Fig.~\ref{Fig:Dephasing}a, with the increase of qubit sensitivity to voltage fluctuation, the $T_2^*$ of Q1 decreases to 1.93 \textmu s when biased at $|\partial f_{\rm{q}}/\partial V_{\rm{res}}| = 184.7$\,MHz/mV, corresponding to 15.9\,MHz detuning from the charge sweet spot. Further, a single refocusing pulse, i.e. Hahn echo, is insufficient to make the coherence time approach the relaxation limit, indicating considerable high-frequency noise components. 

To probe the high-frequency components of the environmental noise on solid neon, we drove the qubit with Carr-Purcell-Meiboom-Gill (CPMG) sequences. 
Qualitatively, the decoherence time under CPMG sequences $T^{\text{CPMG}}_2$ at all the bias points increases with the refocusing number $N$, approaching $2T_1$, as shown in Fig.~\ref{Fig:Dephasing}a.
The extracted pure dephasing time $T_{\phi}$ in Fig.~\ref{Fig:Dephasing}b reflects the frequency-dependent noise power distribution. 
For a noise spectrum following $S(f)~\propto~1/f^{\alpha}$, $T_{\phi}$ scales with $N^{\beta}$, where $\beta = \alpha/(1+\alpha)$ (ref.~\cite{medford2012scaling, jock2022silicon,connors2022charge, yoneda2018quantum}). 
The fitted $\beta$ via this scaling relation provides a more accurate noise distribution compared to individual fittings of the decay curve according to qubit dephasing $\chi_N(\tau) = (\tau/T_\phi)^{\alpha+1}$ (ref.~\cite{medford2012scaling}). 
For Q1, the average fitting gives $\beta$ = 0.6(1), corresponding to $\alpha$ = 1.5(6).

The CPMG sequences act as filter windows in the frequency domain to probe the noise spectral distribution~\cite{bylander2011noise}.
In the case of Q1, the sensing range spans from 0.01 to 1\,MHz, shown as the extracted total noise density in Fig.~\ref{Fig:Noise}a.
(See Methods for details of the CPMG sequence and noise calculation.) 
Figure~\ref{Fig:Noise}b shows the converted voltage (charge) noise on the resonator electrode $S_\text{v}$, scaling from $10^{-4}$ to $10^{-6}$\,\textmu V$^2/$Hz, with a power-law fitting of $S_\text{v} \propto 1/f^{1.3(3)}$. 
A separate measurement on Q2 gives a similar voltage noise density with $S_\text{v}~\propto~1/f^{1.1(3)}$ (Supplementary Information Section 4).
The drop in coherence and the spectral noise distribution observed away from the sweet spot suggest that charge noise dominates the total noise when the qubit is sensitive to electrical fluctuations.

\begin{table*}[hbt]
	\centerline{
		\begin{tabular}{|l|l|l|l|l|l|l|l|}
			\hline
			Qubit platform&Host interface & \begin{tabular}{c} Voltage noise $S_\text{v}$ ($\mu \text{V}^2/$Hz) \\ at $0.01\sim1$ MHz \end{tabular} & Reference & Prospective sources of charge dephasing noise \\ \hline
            
			eNe&Vacuum/Neon  & $10^{-4}\sim10^{-6}$  & This work & Excess electrons on neon;\\  
            
            &  &  & & Substrate surfaces under neon; ref.~\cite{zheng2025surface, kanai2024single}\\ \hline

			Si-MOS&Si/SiO$_2$ & $10^{~0}\sim10^{-2}$  &  ref.~\cite{jock2022silicon} & Surfaces and interfaces of electrical gates;\\ 
   
			&Si/SiO$_2$ & $10^{-1}\sim10^{-3}$  & ref.~\cite{klemt2023electrical} & Si/SiO$_2$ interface; SiO$_2$ layer;\\ 

            &$^{28}$Si/$^{28}$SiO$_2$ & $10^{-1}\sim10^{-3}$  &ref.~\cite{zwerver2022qubits} & ref.~\cite{elsayed2024low, zwerver2022qubits} \\ \hline
            
			Group IV &Si/SiGe & $10^{-3}\sim10^{-5}$  &ref.~\cite{connors2022charge}  & Amorphous gate oxide;\\ 
            
            heterostructures&Si/SiGe & $10^{-1}\sim10^{-3}$ & ref.~\cite{kawakami2016gate} & Strained heterostructure interfaces;\\

            &$^{28}$Si/SiGe  & $10^{-2}\sim10^{-4}$ & ref.~\cite{eng2015isotopically} & In quantum wells confining the electrons;\\
            
			&$^{28}$Si/SiGe & $10^{-2}\sim10^{-4}$  & ref.~\cite{yoneda2018quantum}  & ref.~\cite{paquelet2023reducing, connors2022charge, stehouwer2024exploiting}\\  
            
			&$^{28}$Si/SiGe  & $10^{-1}\sim10^{-3}$  &ref.~\cite{struck2020low} & \\ 
             \cline{2-4}

            &Ge/SiGe  & $10^{-1}\sim10^{-3}$ & ref.~\cite{hendrickx2021four} &\\

            &Ge/SiGe  & $10^{-3}\sim10^{-5}$ & ref.~\cite{stehouwer2024exploiting} & \\ 

            &Ge/SiGe  & $10^{-1}\sim10^{-3}$ & ref.~\cite{hendrickx2024sweet} & \\ \hline

            Group III-V &GaAs/AlGaAs & $10^{-3}\sim10^{-5}$ & ref.~\cite{cerfontaine2016feedback} & Carrier traps in the heterostructure;  \\

			heterostructures&GaAs/AlGaAs & $10^{-6}\sim10^{-7}$ & ref.~\cite{dial2013charge,cerfontaine2020closed} & Gate leakage; ref.~\cite{sakamoto1995distributions, buizert2008n, bermeister2014charge}\\ 
            
            \hline

	  \end{tabular}
            }
	\caption{Comparison of voltage noises and prospective noise origins for electron (hole) qubits on solid neon and in various semiconductor platforms.}
	\label{tab:Charge_noise_table}
\end{table*}

To benchmark the environmental charge noise experienced by electron qubits, Table 1 compares the converted voltage noise on adjacent gate electrodes, located approximately one hundred nanometers from the qubits, across various platforms. The analysis focuses on the voltage noise spectrum from 0.01 to 1 MHz, obtained from characterizations of the eNe charge qubits in this work and the state-of-the-art noise records in silicon metal-oxide-semiconductor (Si-MOS), group IV heterostructures, and group III-V heterostructures.
The high-frequency (between 0.01 and 1 MHz) voltage noise measured at the vacuum/neon interface is in the range between $10^{-4}$ to $10^{-6}~\mathrm{\mu V^2/Hz}$. This voltage noise is at least one order of magnitude lower than in some engineered semiconductor platforms~\cite{eng2015isotopically,jock2022silicon, klemt2023electrical, zwerver2022qubits,yoneda2018quantum, struck2020low, kawakami2016gate, hendrickx2021four, hendrickx2024sweet}
and comparable to others~\cite{connors2022charge, cerfontaine2016feedback, stehouwer2024exploiting}.
This result also approaches some of the best noise data ($10^{-6}$ to $10^{-7}~\mathrm{\mu V^2/Hz}$ in the same frequency range), that have ever been achieved on GaAs/AlGaAs semiconductor platforms~\cite{dial2013charge, cerfontaine2020closed}.
Based on the upper bound of the electrochemical potential lever-arm against a nearby gate (1~eV/V), the estimated charge noise ($\mathrm{10^{-8}}$ to $\mathrm{10^{-9}}$ eV/$\mathrm{\sqrt{Hz}}$ between 0.01 to 1 MHz) is also comparable to the best records in semiconductor qubits~\cite{struck2020low, connors2022charge, stehouwer2024exploiting,elsayed2024low, paquelet2023reducing}.

Among the semiconductor spin qubits used for comparison, GaAs/AlGaAs systems hold the record for the lowest voltage noise~\cite{dial2013charge, cerfontaine2020closed}. This characteristic may be attributed to the high-quality material growth with nearly perfect lattice matching and extremely low disorder~\cite{burkard2023semiconductor}. Nonetheless, the coherence of GaAs/AlGaAs qubits is practically limited by relaxation processes stemming from strong piezoelectric electron-phonon coupling for charge qubits~\cite{stavrou2005charge}, and by nuclear spins for spin qubits~\cite{malinowski2017notch}, making GaAs/AlGaAs not a realistic contender for either charge or spin qubits.

For Si-MOS and group IV based qubits, the origins of the dephasing charge noise are complicated. As summarized in Table 1, these include fluctuators in the bulk material as well as device- and fabrication-dependent interfaces~\cite{paquelet2023reducing,elsayed2024low}. In $\mathrm{Si/SiO_2}$ systems, the extreme proximity of the qubit to the oxide interface and electrical gates can introduce notable disorder and noise~\cite{elsayed2024low, zwerver2022qubits}. In Si/SiGe and Ge/SiGe systems, amorphous gate oxides are known to host a considerable variety of two-level fluctuators~\cite{paquelet2023reducing, connors2022charge, stehouwer2024exploiting}. Mitigating these noise sources in semiconductor platforms often requires intricate material engineering and interface optimization.

The charge noise for the present eNe charge qubits is probably not intrinsic to the solid neon material, but due to the excess electrons trapped on a rough neon surface~\cite{zheng2025surface, kanai2024single} and the surface fluctuators from device substrates. Improving film smoothness and thickness, as well as refining the electron loading procedure, can mitigate these dominant extrinsic noise sources. Despite the current imperfect implementation, the above noise analysis is consistent with the fact that the coherence of eNe charge qubits is orders of magnitude longer than reported semiconductor charge qubits~\cite{stano2022review}. These results not only show solid neon as a low-noise environment for electron charge qubits but also predict a long coherence time for electron spin qubits in this system (Supplementary Information Section 6).

In addition to the high-frequency range, we also investigated the low-frequency noise at the charge sweet spot.
Figure~\ref{Fig:Dephasing}c-d show variations in Q1's frequency revealed by performing repeated Ramsey measurements~\cite{connors2022charge, jock2022silicon}  for approximately one hour since initially biased near its charge sweet spot. 
The qubit undergoes a discrete frequency transition of varying magnitudes, as seen near the 25 minutes of the one-hour tracing. This is revealed by a Fast Fourier transform (FFT) of the Ramsey signals (Methods and Supplementary Information Section 3). Using a periodogram method, we convert this one-hour qubit frequency measurement into a frequency noise spectrum over $10^{-3}$ to $10^{-1}$~Hz, as shown in Fig.~\ref{Fig:Noise}a. A power-law fit of the data gives the relation of $S \propto f^{-1.1(2)}$. Other detection methods, such as single-electron charge sensing techniques~\cite{connors2022charge}, could complete the noise spectrum in the middle-frequency ranging from 1\,Hz to $10^4$\,Hz. 

Although the qubit is first-order insensitive to small charge (voltage) fluctuations at the sweet spot, its local trapping potential remains vulnerable to low-frequency perturbations from second-order effects. These are likely caused by the rearrangement of excess electrons trapped on the rough neon surface, which are difficult to remove because of their low mobility.
Growing smoother neon films and employing gate-controlled electron loading schemes~\cite{koolstra2019coupling} will help mitigate these low-frequency fluctuations in the future. 

To complete the discussion, we calculate the transverse noise at the qubit frequencies contributing to energy relaxation with $1/T_1 = \pi/2\times S(2\pi f_{\rm{q}})$ (ref.~\cite{bylander2011noise}). 
In Fig.~\ref{Fig:Q1_performance}d, the measured $T_1~=~11.6$ \textmu s at Q1's charge sweet spot, where the Purcell rate of Q1 to the resonator mode is $\Gamma_{\rm{r}}~=~\kappa g^2/\Delta_{\rm{rq}}^2~=~1/3.9~\text{ms}^{-1}$. 
$\Delta_{\rm{rq}}$ is the detuning between the resonator and the qubit. And $\kappa$ is the resonator decay rate.
Unlike the qubits reported in our previous work~\cite{zhou2024electron}, this result indicates that non-radiative decay channels dominate the energy relaxation of Q1, as is the case for the other two qubits as well.
Relaxation via phonon emission~\cite{schuster2010proposal, dykman2003qubits, chen2022electron} and interaction with excess surface electrons, which can act as background bath of two-level systems~\cite{muller2015interacting}, may limit the relaxation lifetime of floating electron qubits.
Measurements of Q1's $T_1$ at various bias frequencies show that transverse noise increases with frequency in GHz range, probably due to the interaction with an ohmic-type environment~\cite{paladino20141}.  
The best fit of the total transverse noise versus frequency in log scale gives $S \propto f^{2.2(2)}$ between 5.065 to 5.498~GHz, with a mean of $3.6\times 10^5$~$\text{Hz}^2/\text{Hz}$, as shown in Fig.~\ref{Fig:Noise}c.
\\

\noindent\textbf{Temperature dependence}

\begin{figure*}[htb]
\centerline{\includegraphics[scale=0.211]{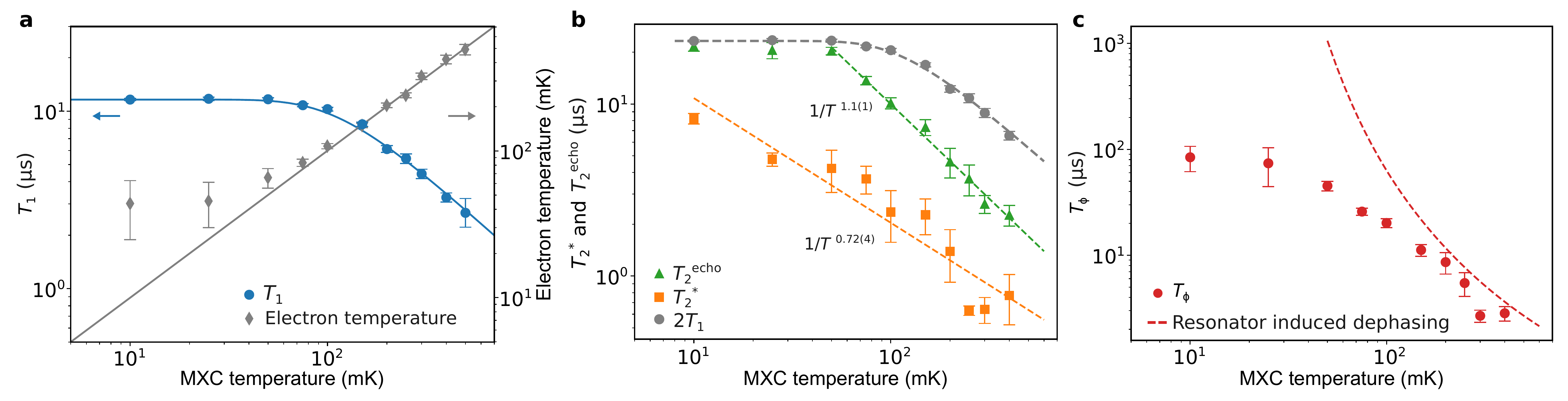}}
\caption{\textbf{Temperature-dependent coherence of eNe qubit Q1 at charge sweet spot}. \textbf{a}, Relaxation time $T_1$ (blue dots, data) versus mixing chamber (MXC) temperature. The solid curve represents the predicted $T_1(T) = T_1(T=0)\cdot\text{tanh} (\hbar \omega_{\rm{q}}/2k_BT)$ where we use the measured value at 10\,mK for $T_1(T=0)$, and the MXC temperature as $T$. Extracted electron temperature (gray diamonds, data) versus MXC temperature based on the measured thermal population in Supplementary Information Section 3. \textbf{b}, Decoherence time, $T_2^*$ (orange squares) and $T_2^{\rm{echo}}$ (green triangles) versus the MXC temperature, with power-law fitting (dashed curves). Gray dots and curves show $2T_1$ for comparison. \textbf{c}, Extracted pure dephasing time $T_{\phi}$ as a function of the MXC temperature. Red dashed curve: Parameter-free calculation of the resonator-induced dephasing based on Q1 properties. All error bars represent the one standard error of extracted parameters (Methods).} \label{Fig:Temperature}
\end{figure*}

To characterize the robustness of solid neon film as electron qubit host at elevated temperatures, we measured the temperature dependence of Q1's relaxation and coherence at its charge sweet spot from 10\,mK to 500\,mK.
The energy relaxation $T_1$ is well described by a model that only accounts for the coupling of a two-level quantum system to a bosonic thermal bath within the experiment's temperature range, $T_1\propto\text{tanh} (\hbar \omega_{\rm{q}}/2k_{\text{B}}T)$ (ref.~\cite{ lisenfeld2010measuring, leggett1987dynamics}), as shown in Fig.~\ref{Fig:Temperature}a. Here, $\hbar$ is the reduced Planck constant, $k_{\text{B}}$ is the Boltzman constant, and $T$ is temperature.
Similarly, the thermal population follows that of a Maxwell-Boltzmann distribution~\cite{jin2015thermal} (Supplementary Information Section 3).
The corresponding electron temperature closely tracks the mixing chamber (MXC) temperature between 100\,mK to 500\,mK and saturates around 40~mK, showing effective cooling of electrons at low temperatures.

A separate measurement on a different qubit Q2, presented in Supplementary Information Section 4, reveals a similar temperature dependence in $T_1$ but with a different low-temperature limit.
It suggests a picture in which individual eNe qubits couple differently to their environment (including through phonons~\cite{schuster2010proposal, chen2022electron}), which sets the low temperature $T_1$.
This motivates further studies of the microscopic limitations of $T_1$.
For the current qubits with transition frequencies near 5\,GHz, $T_1$ decreases to about half of its low-temperature value at 200~mK. 
Given the temperature-dependent $T_1$ scaling with the two-level quantum system model, we may anticipate further improvements of the high-temperature performance of eNe qubits by engineering charge or spin states for qubit operations at higher frequencies. 
This would motivate the development of dedicated cQED architectures and control hardware,
for which recent efforts have already begun in superconducting qubits operating above 20 GHz~\cite{anferov2024superconducting}.

The coherence data at elevated temperatures (Fig.~\ref{Fig:Temperature}b) shows solid neon's thermal robustness as an electron qubit host.
Q1 with qubit frequency only at $~\sim$5\,GHz maintains $>$~1~\textmu s $T_2^{\rm{echo}}$ at temperatures up to 400~mK, suggesting eNe's potential for high-temperature operation. 
Meanwhile, the trend of coherence with varying temperature remains complex.
$T_2^*$ and $T_2^{\rm{echo}}$ show that quasi-static and high-frequency noise components behave differently with increased temperature, as shown in Fig.~\ref{Fig:Temperature}b.
Below 50~mK, $T_2^{\rm{echo}}$ approaches $2T_1$, suggesting a weak noise density in the high-frequency range.
However, starting from 75~mK, $T_2^{\rm{echo}}$ scales with temperature as $\propto T^{-1.1(1)}$, and quickly degrades from $2T_1$, indicating the rise of non-quasi-static dephasing noises at higher temperatures~\cite{dial2013charge}.
On the other hand, $T_2^* \propto T^{-0.72(4)}$ in the whole testing range.

Extracted pure dephasing time ($T_{\phi}$) indicates qubit's decoherence due to non-quasi-static noise grows primarily due to thermal effects at elevated temperatures.
As shown in Fig.~\ref{Fig:Temperature}c, $T_{\phi}$ begins to decrease notably when the MXC temperature exceeds 100~mK.
In this range, the measured data matches well with a parameter-free model accounting for the effect of resonator thermal photons on the qubit dephasing~\cite{anferov2024superconducting}:
\begin{equation}
T_{\phi}^{-1} = \frac{\kappa}{2} \text{Re} \left[\sqrt{\left(1+\frac{2i\chi}{\kappa}\right)^2+\frac{8i\chi}{\kappa}n_{\text{th}}}-1\right]
\end{equation}
where $\chi$ is the resonator's dispersive shift (Supplementary Information Section 3).
$n_{\text{th}}$ represents the resonator thermal photon population, equal to $1/(e^{hf_{\rm{r}}/k_{\text{B}}T}-1)$, with $f_r$ as the resonator frequency.
Nevertheless, the data's deviation from the model below 100\,mK suggests the presence of other low-temperature dephasing mechanisms.
Separate measurements on Q2 (Supplementary Information Section 4) give similar results. Using higher frequency resonators~\cite{anferov2024superconducting} can help to suppress the photon-noise induced dephasing.\\

\noindent\textbf{Conclusions}

We have reported a quantitative investigation of the environmental noise isolation provided by the thin neon layer for eNe qubits. 
We have shown the eNe qubits are resilient to temperature and could be used for high-temperature operation. 
Improving the system’s consistency in terms of qubit properties and noise isolation will be an important next step in the development of eNe qubits. 
Our work also highlights the importance of improving neon growth, as well as electron generating and trapping techniques.

The varied spectral properties of eNe qubits indicate that the local neon profile is crucial in electron-trapping mechanisms. 
The surface roughness of cryogenic solid neon and hydrogen films has been studied using optical methods and surface electron mobility measurements~\cite{leiderer2025surface}. 
With the current setup and neon filling approach, the cryogenic adsorption of neon back into 4K section of the fill lines during the cool-down, as well as the triple-point (de)wetting~\cite{leiderer2025surface} may result in a large surface roughness on the thin neon films.
The morphology of the underlying substrate could also translate to the neon surface and affect the properties of trapped eNe qubits~\cite{zheng2025surface}.
Developing neon growth method with more refined control~\textemdash~including cryogenic valve~\cite{dotsenko2004really} and temperature regulation, and repetitive annealing~\cite{matkovic2025characterizing}~\textemdash~as well as exploring condensation methods~\cite{kanagin2025impurities}, could improve neon film quality and qubit uniformity.
Additionally, the relative position of trapped electrons to nearby electrodes determines the frequency lever-arm and the qubit’s sensitivity to voltage fluctuations. 
Currently, the lack of a precise method for growing solid neon makes the electron trapping uncertain, which induces varied qubit properties.

Improving the consistency and control over individual eNe qubits is also important for scaling up. 
All-microwave and fast DC types of two-qubit gates could be applied to eNe qubits coupled through resonator bus~\cite{majer2007coupling} or direct charge–charge interactions~\cite{beysengulov2024coulomb}. 
They all require stable and efficient control over qubit properties. 
In both regards, refined neon growth methods, gated electron loading mechanisms, and electron detectors with higher spatial resolution should be developed.

On the other hand, diverse noise behaviour observed in eNe qubits reflects the complexity of the local charge environment.
We have shown that for two eNe qubits, a single refocusing pulse is insufficient to mitigate the majority of noises at the qubits’ sweet spot, suggesting a higher noise density at the high-frequency range (Supplementary Information Section 4 and 5).
The different spectral noise distributions suggest locally non-uniform noise sources, as qubit’s coherence is highly sensitive to fluctuators’ density and distribution~\cite{mehmandoost2024decoherence}.
Excess electrons emitted by the tungsten filament could result in nonuniformly distributed charge fluctuators on the neon surface. 
Rearrangement of excess charges adjacent to the qubit may contribute to the low-frequency qubit drift and occasional electron escape events~\cite{kuhlmann2013charge}.
Fluctuation of distant charges may also perturb the qubits via the high-impedance resonator due to the enhanced coupling strength. 
The thin neon film thickness could further reduce the effectiveness of isolating noises embedded in the underlying substrate.
Strategies to improve the qubit’s coherence include developing electron loading procedures on thicker  ($\sim$100\,nm) films~\cite{koolstra2019coupling, tian2025nbtin} and separating trapped qubits from adjacent charge reservoirs~\cite{yang2020operation} with protection gates.

Finally, the limiting factor of the non-radiative energy relaxation rate of eNe qubits needs further investigation. 
The non-monotonic variations in $T_1$ of qubits with different charge sweet spot frequencies point to varied non-radiative decay channels.
Nearby charges may create a sparse bath of two-level fluctuators weakly coupled to the eNe qubit, whose density determines the transverse noise intensity~\cite{mehmandoost2024decoherence}.
Electrical gating could be applied to bias or repel those weakly coupled charges to reduce the relaxation rate~\cite{zheng2022coherence}. 
In addition, the rough neon surface may create more channels for phonon-induced relaxation, compared with electrons on flat neon or helium, due to breaking the symmetry of otherwise forbidden selection rule processes~\cite{dykman2003qubits, chen2022electron}. 
With the development of improved electron generating and trapping methods on thicker and smoother neon, we anticipate improvements in both the coherence and long-term stability of eNe qubits. \\

\noindent\textbf{Methods}

\noindent\textbf{Device and setup}

The resonator, electrode, and on-chip filter were patterned with electron beam lithography followed by reactive-ion etching. Around the traps, the Si substrate is etched down by approximately 250~nm to host the thin neon layer. Unlike the previous device, originally designed for trapping electrons on helium~\cite{zhou2022single, zhou2024electron}, where the resonator is embedded in a 1\,\textmu m-deep channel to host liquid helium, the TiN resonator in our device is positioned at the same level as the ground plane~\cite{zheng2025surface}. Besides, the two-trap design is for the broader goal of coupling two distant eNe qubits via the resonator bus~\cite{majer2007coupling}. See details of the TiN film, resonator, and on-chip filter characterization in Supplementary Information Section 1. 

The experiment setup is similar to the ones in our previous works~\cite{zhou2022single, zhou2024electron}. The chip was mounted on a customized printed circuit board within a vacuum-sealed copper cell, which provides direct current (DC) and microwave (MW) interfaces. On the top lid of the cell, a gas filling line was attached to deposit neon at cryogenic temperature, and a tungsten filament was used as the electron source. The cell is mounted on the mixing chamber (MXC) plate of a dilution refrigerator. A total attenuation of 60\,dB was applied on the cryogenic segment of the MW input line with infrared filters (QMC-CRYOIRF-004). The MW output line was equipped with cryogenic isolators (LNF-ISISC4\_12A) at the MXC plate, followed by a high electron mobility transistor amplifier (LNF-LNC4\_8C) at 4\,K plate and room temperature amplification. All DC connections were filtered with thermocoax cables, inductor-capacitor (LC) filters (Mini-Circuits RLP-30+), and home-made low-pass filters with 150 Hz cut-off. Qubit spectroscopic measurements were conducted with a vector network analyser (Keysight N5222B) and a signal generator (Anritsu MG3692C). Time-domain pulse measurements were conducted with Quantum Machine OPX+ and Octave. The DC gates were applied with QDevil QDAC-II.\\

\noindent\textbf{Neon growth}

Neon is filled with the following procedure. The fridge is warmed up from its base temperature with a heater mounted on the 4~K plate. At this moment, the helium mixture circulation is turned off, and all the mixture has been collected while the pulse tube is still on. The heater power is set to a value such that it creates a temperature gradient from 27~K at the 4~K plate to about 25~K at the MXC plate. Under such conditions, the neon gas is filled and deposited onto the device chip in its liquid phase. After filling, the heater is turned off, and the whole fridge is cooled down again to let the liquid neon freeze into solid. During the cool down, we further anneal the neon film at 10~K for 1 hour. \\

\noindent\textbf{Electron deposition}

Electrons are ejected from the tungsten filament mounted on the lid of the hermetic copper cell. When the dilution fridge is cooled down to the base temperature, the total resistance on the filament loop is 2~$\Omega$. A current pulse train applied by a pulse generator with $-0.6$~V voltage output, pulse width of 0.1~ms, repetition frequency of 1~kHz, and duration of 0.3~s was used to fire electrons. We noticed that applying higher voltage or longer duration would cause too many electrons to land on the top of the chip, reducing the stability of the trapped electrons. Only when both the neon film is present on the sample surface and electrons are emitted do we see signatures of trapped eNe qubits strongly coupled to the resonator (Supplementary Information Section 7).\\

\noindent\textbf{Noise characterization}

We use the dynamical decoupling technique with Carr-Purcell-Meiboom-Gill (CPMG) pulse sequences to study the spectral distribution of high-frequency noise affecting the eNe qubits~\cite{bylander2011noise}. All qubit bias points had positive $\Delta V_{\rm{res}}$. In the sequence, $N$ refocusing $Y_\pi$ pulses are applied between two $X_{\pi/2}$ pulses with identical separation $\tau/N$ between two pulses. Under such sequences, qubits' coherence decaying follows
$P_\text{e}(N,\tau)=P_0+a\cdot\text{exp}\left(-\tau/2T_1\right)\text{exp}\left(-\chi_\text{P}\right)\text{exp}\left(-\chi_N(\tau)\right)$, accounting for pure dephasing $-\chi_N$, energy relaxation $-\tau/2T_1$, and decay during driving pulses $-\chi_\text{P}$ (ref.~\cite{bylander2011noise}). We define $T_2^{\text{CPMG}}$ as the time when the qubit's coherence decays by a factor of $1/e$ due to energy relaxation and pure dephasing. Increasing the number of $Y_\pi$ pulses reduces the time the qubit is exposed to noise before each refocusing, thereby extending the coherence time. Under Gaussian noise assumption~\cite{bylander2011noise}, a CPMG sequence with $N$ refocusing pulses and total delay time of $\tau$ imposes a filter function $g_N(\omega,\tau)$ to the qubit noise caused by source $\lambda$ with spectral distribution $S_\lambda(\omega)$, which determines the qubit's dephasing $\chi_N(\tau)$:
\begin{equation}
 \chi_N(\tau)=\tau^2\sum_{\lambda}^{} \left(\frac{\partial \omega_\text{q}}{\partial \lambda}\right)^2 \int_{0}^{\infty} S_\lambda(\omega)g_N(\omega,\tau) \,d\omega 
\end{equation} where $\partial\omega_{\rm{q}}/\partial\lambda$ is the qubit transition frequency's sensitivity to noise source $\lambda$, the filter function 
$
g_N(\omega,\tau)=|y_N(\omega,\tau)|^2/(\omega\tau)^2
$, and
$y_N(\omega,\tau) = 1+(-1)^{1+N}\text{exp}(i\omega\tau)$
$+ 2\sum_{j=1}^{N}(-1)^{j}\text{exp}(i\omega\tau(j-0.5)/N)\text{cos}(\omega\tau_{\pi}/2)
$. 

With the noise filter imposed by the CPMG sequence, we could approximate the total noise power spectrum density $S$ at frequency $f_N$ as (ref.~\cite{connors2022charge,jock2022silicon}):
\begin{equation}
S(2\pi f_N) = \frac{1}{(T_\phi)^2g_N(2\pi f_N, T_{\phi}) \Delta\omega_N}
\end{equation}
where $2\pi f_N$ is the peak angular frequency of the first harmonic of $g_N(\omega, T_\phi)$, and $\Delta \omega_N$ is the full width at half maximum of the peak. 
$T_\phi$ is the time when $\chi_N(T_\phi)=1$.
Given the qubit's sensitivity $|\partial f_\text{q} / \partial V_\text{res}|$, as listed in the legends of Fig.\,3 and 4, and neglecting other potential noise sources, we extract the equivalent voltage noise ($S_\text{v}$) as:
\begin{equation}
S(2\pi f) = (\partial f_\text{q} / \partial V_\text{res})^2 \times S_\text{v}(2\pi f)
\end{equation}

To investigate the low-frequency dynamics of qubit decoherence near the charge sweet spot, we repeatedly conducted Ramsey measurements for 128 iterations, with each recording taking 33 seconds~\cite{connors2022charge,jock2022silicon}. As shown in Fig.~\ref{Fig:Dephasing}d, we extracted the qubit frequency drift through a fast Fourier transform (FFT) analysis of the measured Ramsey fringes. During the measurement, we also observed more drastic qubit frequency fluctuations on the scale of 10~--~100~MHz, corresponding to 0.14~--~0.44~mV in $\Delta V_{\text{res}}$. Sometimes, we even observed the disappearance of qubit signatures in the usual gate scanning range. These qubit frequency ``jumps" usually occur less than 0.5 times per hour. We attribute these large fluctuations to charge rearrangements of nearby weakly trapped electrons on neon~\cite{kuhlmann2013charge}. We ensured that the qubit remained relatively stable and free from such dramatic fluctuations during data collection for all the presented measurements at various bias points.

For all time-domain decoherence and relaxation measurements, each data point is averaged over 500 times of single-shot readouts.\\

\noindent\textbf{Data availability}
The source data of this study are available via Zenodo at https://doi.org/10.5281/zenodo.18548730. The rest data that support the findings are available from the corresponding authors upon request. Source data are provided with this paper.\\

\noindent\textbf{Code availability}

The codes used for analysing and plotting the data in this work are available from the corresponding authors upon request.\\

\acknowledgements

D. J., X. H., and X. L. acknowledge support from the Argonne National Laboratory Directed Research and Development (LDRD) program for experimental setup. D. J., X. L., and Y. W. acknowledge support from the Air Force Office of Scientific Research (AFOSR) under Award No. FA9550-23-1-0636 for device fabrication and simulation. D. J. acknowledges support from the Department of Energy (DOE) under Award No. DE-SC0025542 for material growth and characterization. D. J. and W. G. acknowledge support from the National Science Foundation (NSF) under Award No. OSI-2426768 for theoretical modelling. D. J., X. Zhou, and Y. H. acknowledge support from the Julian Schwinger Foundation for Physics Research for instrument development. B. D. acknowledges support from the NSF under Award No. DMR-1906003. C. S. W. and X. L. acknowledge support from Q-NEXT, one of the US DOE Office of Science National Quantum Information Science Research Centers. W. G. acknowledges support from the Gordon and Betty Moore Foundation through Grant DOI 10.37807/gbmf11567 and the National High Magnetic Field Laboratory at Florida State University, which is supported by the NSF Cooperative Agreement No. DMR-2128556 and the state of Florida. X. Zhang and Y. H. acknowledge support from the Office of Naval Research (ONR) Young Investigator Program (YIP) program under Award No. N00014-23-1-2144. X. H. acknowledges support from France and Chicago Collaborating in the Sciences (FACCTS) program. Work performed at the Center for Nanoscale Materials, a U.S. Department of Energy Office of Science User Facility, was supported by the U.S. DOE, Office of Basic Energy Sciences, under Contract No. DEAC02-06CH11357. The authors thank Xuedong Hu, Robert Joynt, David I. Schuster and Amir Yacoby for helpful discussions.\\

\noindent\textbf{Author contributions}
X. L., X. Zhou, and D. J. devised the experiment. X. H. and D. J. advised the experiment. X. L. and C. S. W. conducted the experiment. X. L. designed and fabricated the device. X. L., C. S. W., and B. D. analysed the data. Y. H., X. Zhang, and X. H. supported the experimental measurement. Y. W. and W. G. supported the theoretical calculation. D. J. conceived the idea and led the project. X. L., C. S. W., X. Zhou, and D. J. wrote the original manuscript. All authors contributed to the work.\\

\noindent\textbf{Competing Interests Statement}
The authors declare no competing interests.

\bibliography{eNe_noise}

\clearpage

\end{document}


\title{Supplemental Information for\\ \centerline{Solid neon as a noise-resilient host for electron qubits above 100 mK}}

\author{Xinhao Li}
\affiliation{Center for Nanoscale Materials, Argonne National Laboratory, Lemont, Illinois 60439, USA\looseness=-1}
\affiliation{Department of Physics, Harvard University, Cambridge, Massachusetts 02138, USA\looseness=-1}

\author{Christopher S. Wang}
\affiliation{James Franck Institute and Department of Physics, University of Chicago, Chicago, Illinois 60637, USA\looseness=-1}

\author{Brennan Dizdar}
\affiliation{James Franck Institute and Department of Physics, University of Chicago, Chicago, Illinois 60637, USA\looseness=-1}

\author{Yizhong Huang}
\affiliation{Center for Nanoscale Materials, Argonne National Laboratory, Lemont, Illinois 60439, USA\looseness=-1}
\affiliation{Department of Electrical and Computer Engineering, Northeastern University, Boston, Massachusetts 02115, USA\looseness=-1}

\author{Yutian Wen}
\affiliation{Department of Physics and Astronomy, University of Notre Dame, Notre Dame, Indiana 46556, USA\looseness=-1}

\author{Wei Guo}
\affiliation{National High Magnetic Field Laboratory, Tallahassee, Florida 32310, USA}
\affiliation{Department of Mechanical Engineering, FAMU-FSU College of Engineering, Florida State University, Tallahassee, Florida 32310, USA}

\author{Xufeng Zhang}
\affiliation{Department of Electrical and Computer Engineering, Northeastern University, Boston, Massachusetts 02115, USA\looseness=-1}

\author{Xu Han}\email[Email: ]{xu.han@anl.gov}
\affiliation{Center for Nanoscale Materials, Argonne National Laboratory, Lemont, Illinois 60439, USA\looseness=-1}
\affiliation{Pritzker School of Molecular Engineering, University of Chicago, Chicago, Illinois 60637, USA\looseness=-1}

\author{Xianjing Zhou}\email[Email: ]{xianjing.zhou@fsu.edu}
\affiliation{National High Magnetic Field Laboratory, Tallahassee, Florida 32310, USA}
\affiliation{Department of Mechanical Engineering, FAMU-FSU College of Engineering, Florida State University, Tallahassee, Florida 32310, USA}

\author{Dafei Jin}\email[Email: ]{dfjin@nd.edu}
\affiliation{Center for Nanoscale Materials, Argonne National Laboratory, Lemont, Illinois 60439, USA\looseness=-1}
\affiliation{Department of Physics and Astronomy, University of Notre Dame, Notre Dame, Indiana 46556, USA\looseness=-1}

\date{\today}

\maketitle

\tableofcontents

\section{Titanium nitride resonator}

eNe qubits were coupled with titanium nitride (TiN) high-impedance superconducting resonators. In this section, we describe the details of the resonator device.

\subsection{TiN thin film}
30\,nm TiN thin film was grown on intrinsic $\langle111\rangle$ Silicon via atomic layer deposition following the procedure in ref.~\cite{shearrow2018atomic}. Figure S1a plots the sheet resistance as the function of substrate temperature measured by a physical property measurement system. The kinetic inductance of the film is estimated with (ref.~\cite{shearrow2018atomic}):

\begin{equation}
  L_\text{k} = \hbar R_{\square} / \pi \Delta_{0}
\end{equation}

Where $R_{\square}$ is the normal sheet resistance measured just above $T_c$, and $\Delta_{0}=1.76~k_\text{B}T_\text{c}$ is the superconducting energy gap for TiN. With the measured $T_\text{c} = 3.8~\text{K}$ and $R_{\square} = 64.42~\Omega/\square$, the estimated kinetic inductance $L_\text{k} = 23.4~\text{pH}/\square$. 

\begin{figure}[htb]
	\centering
	\includegraphics[scale=0.4]{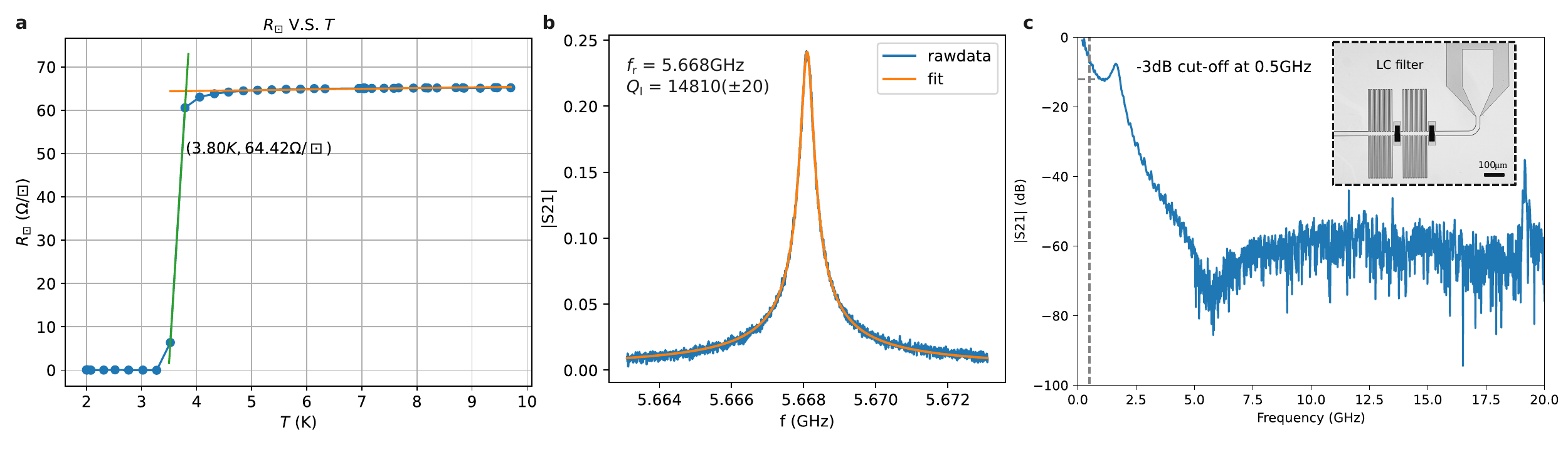}
	\caption{\textbf{Titanium nitride (TiN) thin film and resonator.} \textbf{a}, Sheet resistance $R_{\square}$ of 30\,nm TiN film on intrinsic $\langle111\rangle$ Silicon substrate, measured at various temperature $T$. \textbf{b}, Transmission spectrum $\mathrm{|S21|}$ of the resonator's differential mode, with $f$ denoting the frequency. \textbf{c}, Transmission through the on-chip LC low pass filter. Inset: Microscope photo of the filter.}
	\label{Fig:Resonator}
\end{figure}

\subsection {Resonator and filter}

Figure S1b plots the measured and fitted differential mode at 5.668\,GHz, with loaded quality factor $Q_\text{l} = 14810\,(\pm20)$, or equivalent resonator linewidth $\kappa/2\pi = 0.38$\,MHz\,($\pm~0.5$\,kHz). On a similar device without gate lines, fabricated from the same batch of TiN film, we characterized its quality factors by measuring the transmission and reflection spectra through the input and output ports. The measured loaded quality factor $Q_\text{l} = 1.184 \times 10^4\,(\pm 30)$, $Q_\text{c,in} = 2.44 \times 10^4\,(\pm 3\times10^2)~\text{and}~ Q_\text{c,out} = 2.98 \times 10^4\,(\pm 4\times 10^2)$, yielding the intrinsic quality factor $Q_\text{i} = 1.01 \times 10^5\,(\pm 7\times10^3)$. It indicates the high quality of the TiN film and that the coupling through the input and output ports, as well as the gate lines, limits the loaded quality factor of the device under test. We estimated the resonator's $\lambda/2$ differential mode impedance with the equivalent lumped element model of a split superconducting resonator (ref.~\cite{koolstra2019coupling,koolstra2025high}):

\begin{equation}
  \omega_\text{diff} = \frac{1}{\sqrt{L_\text{diff}C_\text{diff}}}
\end{equation}
\begin{equation}
  L_\text{diff} = \frac{2L_\text{l}l}{\pi^2}
\end{equation}
\begin{equation}
  Z_\text{diff} = \sqrt{\frac{L_\text{diff}}{C_\text{diff}}}
\end{equation}

Where $l = 1061$\,\textmu m is the length of the inductance part of the resonator. For a split resonator, self and mutual inductance contribute to the total inductance as $L_\text{l} = L_\text{self} + |L_\text{mutal}|$. Here, we estimated $L_\text{l} \simeq L_\text{k}/w = 7.8\times10^{-5}$\,H/m, where resonator width $w = 0.3$\,\textmu m, neglecting the mutual inductance due to the large distance between the two resonator pins and the minor geometry contribution to the self-inductance. Following the model, the equivalent $L_\text{diff}~=~1.68\times10^{-8}~\text{H}$ and $C_\text{diff}~=~4.69\times10^{-14}~\text{F}$. The calculated equivalent differential mode impedance is $Z_\text{diff}~=~598.5~\Omega$. Given the qubit-resonator coupling strength $g\propto\omega_\text{r}\sqrt{Z_\text{r}}$ (ref.~\cite{koolstra2025high}), the TiN resonator should improve the coupling strength by about 2.85 times compared with the previous niobium resonator with equivalent lumped element impedance of $\sim57.3\,\Omega$ and differential mode at 6.42\,GHz (ref.~\cite{koolstra2019trapping}).

To maintain the resonator's quality factor, on-chip LC filters are attached to all DC electrodes near the electron trap and to the center bias on the resonator. Figure S1c plots the measured transmission through the LC filter showing broadband $\sim 60~\text{dB}$ isolation above 4\,GHz with a $3~\text{dB}$ cut-off frequency at 0.5\,GHz.

\section{Summary of qubit properties}

This section summarizes the properties of the three eNe qubits tested in this work.

\begin{table}[h]
    \centering
    \begin{tabular}{|l|l|l|l|l|l|l|}
    \hline
     & $f_\text{ss}$ (GHz) & $T_\text{1}$ (\textmu s) & $T_\text{2}^*$ (\textmu s) & $T_\text{2}^\text{echo}$ (\textmu s) &  $N$ to approach 2$T_\text{1}$ & $g/2\pi$ (MHz)  \\ \hline
    Qubit 1 & 5.065 & 11.6 & 8.2 & 21.6 & 2  & 6.43  \\  \hline
    Qubit 2 & 5.699 & 25.1 & 7.2 & 25.2 & 64 & 6.17 \\ \hline
    Qubit 3 & 4.456 & 74.3 & 3.5 & 4.0 & 128 & 16 \\ \hline
    \end{tabular}
    \caption{Summary of the three eNe qubits' properties at charge sweet spot. $f_\mathrm{ss}$: charge sweet-spot frequency; $T_1$: relaxation time; $T_2^*$: total dephasing (Ramsey) time; $T_2^\mathrm{echo}$: total decoherence time with one Hahn echo; $N$: number of echo pulses in dynamical decoupling sequence; $g/2\pi$: qubit-resonator coupling strength. }
    \label{tab:Qubit_table}
\end{table}

\section{Qubit 1 (Q1) supporting data}

\subsection {Qubit spectroscopy}

Figure S2a plots the data and fit of the transmission spectrum when the qubit is in resonance with the resonator. Fitting to the input-output theory model gives the qubit-resonator coupling strength $g/2\pi~=~6.43$\,MHz and qubit linewidth $\gamma/2\pi~=~3.18$\,MHz.

\begin{figure}[htb]
	\centering
	\includegraphics[scale=0.38]{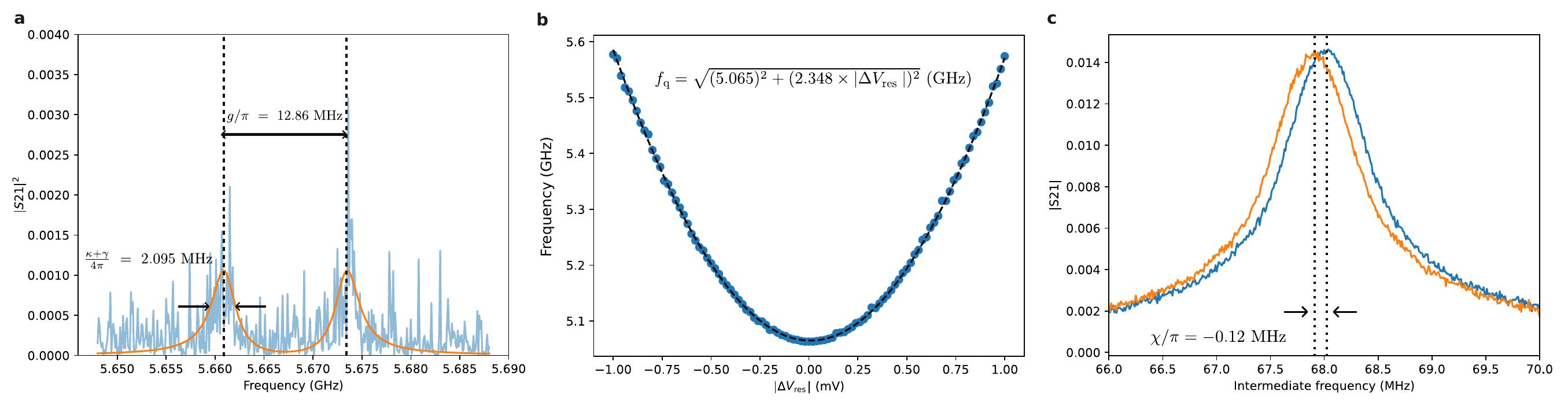}
	\caption{\textbf{Qubit 1 spectroscopy data.} \textbf{a}, Measurement and fitting of transmission spectrum ($\mathrm{|S21|}$) when Q1 is on resonance with the resonator. $g$ represents the qubit-resonator coupling strength. $\kappa$ and $\gamma$ are the resonator and qubit linewidth, respectively. \textbf{b}, Fitting of the qubit spectrum ($f_\mathrm{q}$) as the function of the relative resonator bias voltage $\Delta V_\mathrm{res}$ to the model in the main text. \textbf{c}, Resonator dispersive shift measurement when Q1 is biased at its charge sweet spot.}
	\label{Fig:Q1_spec}
\end{figure}

Figure S2b plots the qubit spectrum with the transition frequency extracted from Fig. 2b in the main text. We fit the qubit spectrum to $f_\text{q} = \sqrt{(2.348\times10^{9}\Delta V_\text{res})^2+(5.065\times10^{9})^2}$, with $f_\text{q}$ in the unit of Hz and $\Delta V_\text{res}$ in the unit of mV. The equivalent $\Delta V_\text{res}$ and qubit sensitivity to voltage fluctuation in the main text is calculated based on this fitting result, given the measured qubit frequency at different basis points. This generalized model has been widely applied to describe two-level quantum systems~\cite{muller2019towards}, with an energy landscape consistent with our bounds of large anharmonicity in eNe charge qubits~\cite{zhou2024electron}.

Figure S2c plots the transmission spectrum through the resonator when the qubit is prepared at the ground state (blue curve) or excited state (orange curve) biased at its charge sweet spot. The measurement indicates a dispersive shift $\chi/\pi~=~-0.12$\,MHz.
\subsection {Qubit frequency drift}

\begin{figure}[htb]
	\centering
	\includegraphics[scale=0.32]{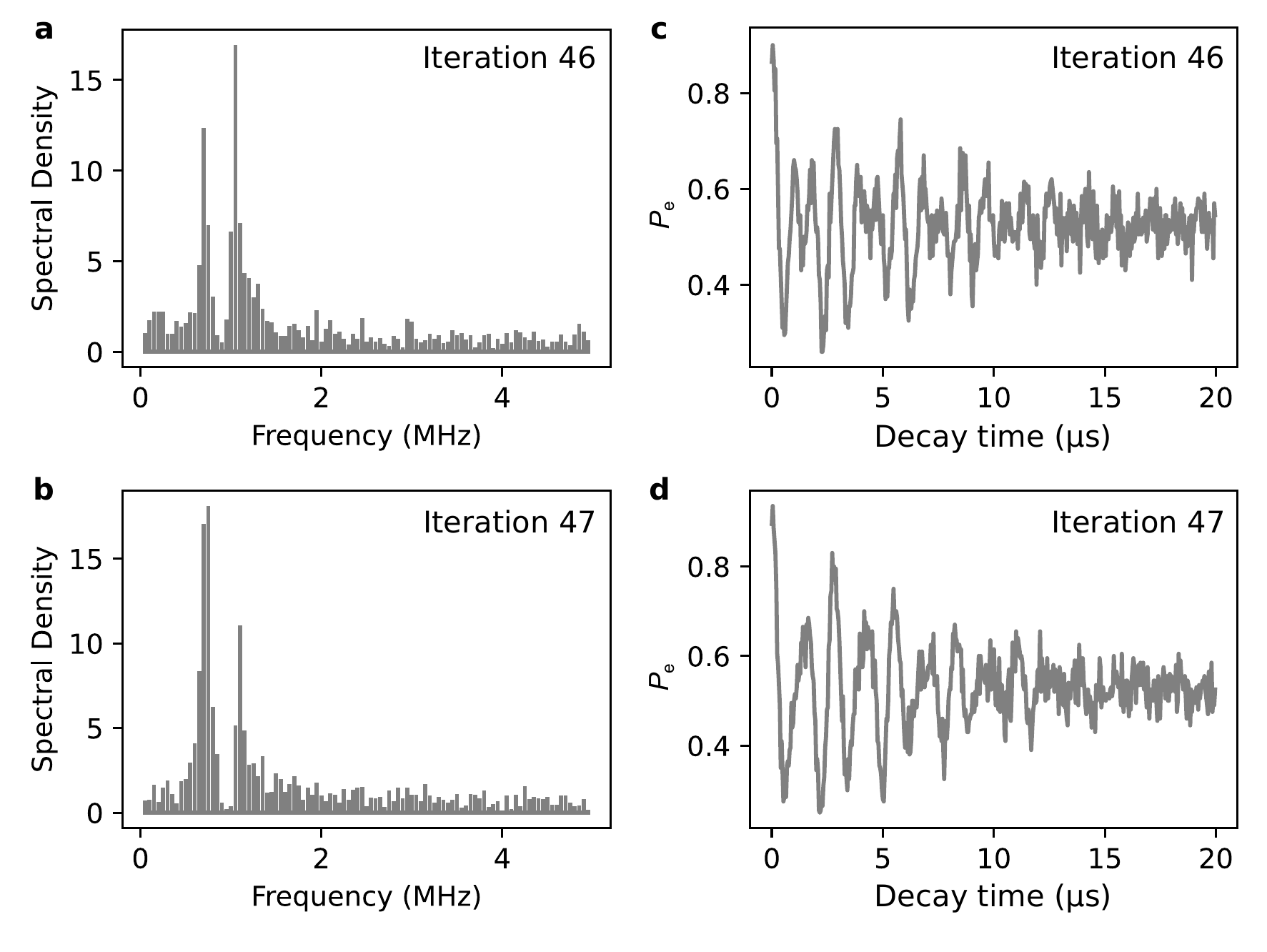}
	\caption{\textbf{Qubit 1 frequency drift.} \textbf{a} and \textbf{b}, Frequency components of Ramsey measurements in iterations 46 and 47. \textbf{c} and \textbf{d}, Raw data of Ramsey measurements in iterations 46 and 47. $P_\mathrm{e}$ represents the qubit's excited state population.}
	\label{Fig:Q1_spec}
\end{figure}

Figure S3 plots the measured Ramsey fringes in iterations 46 and 47, whose frequency components show double peaks. These results reveal a slow drift of the qubit frequency during the one-minute measurement.

\subsection {Thermal population}

Figure S4 plots the measured thermal population of Q1 with standard deviations versus the extracted electron temperature. Following a Maxwell-Boltzmann model~\cite{jin2015thermal}, and considering only the first two qubit states, the thermal-induced qubit's excited state population $P_\text{e} = 1/(1+1/\text{exp}(-\hbar\omega_\text{q}/ k_\text{B}T_\text{e}))$, where $\omega_\text{q}$ is the qubit frequency at the sweet spot, we extracted the electron temperature $T_\text{e}$ and its error bar, as shown in Fig. 5a of the main text.
\begin{figure}[htb]
	\centering
	\includegraphics[scale=0.6]{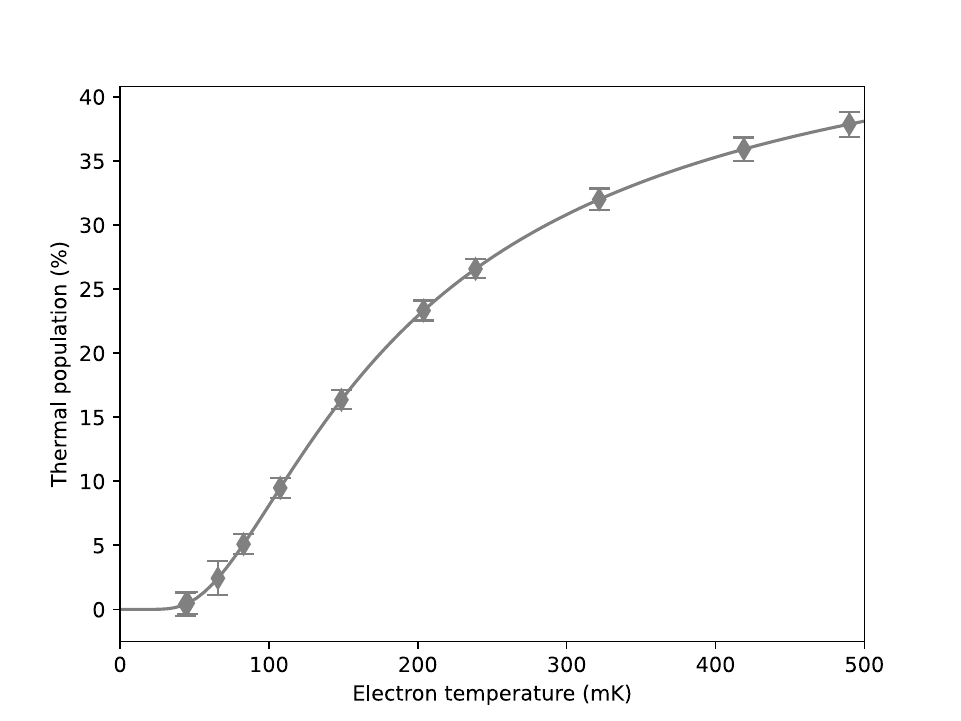}
	\caption{\textbf{Thermal population of Q1 versus electron temperature.} Thermal population of Q1 (gray diamonds, data) along with the predicted population (gray solid curve) from a Maxwell-Boltzmann distribution model with extracted electron temperature. Error bars represent the one standard error of the extracted excited state population from 5000 single-shot measurements.}
	\label{Fig:Q1_spec}
\end{figure}

\section{Qubit 2 (Q2) supporting data}

Q2 was trapped on a different resonator fabricated within the same batch as the one used to trap Q1. The resonator has a differential mode at 5.870\,GHz with a resonator linewidth of $\kappa/2\pi~=~0.53$\,MHz. We performed noise characterization with dynamical decoupling methods when it was biased on and off the charge sweet spot. We also characterized its temperature-dependent coherence performance on the charge sweet spot. The details of the data are shown in this section.

\begin{figure}[htb]
	\centering
	\includegraphics[scale=0.28]{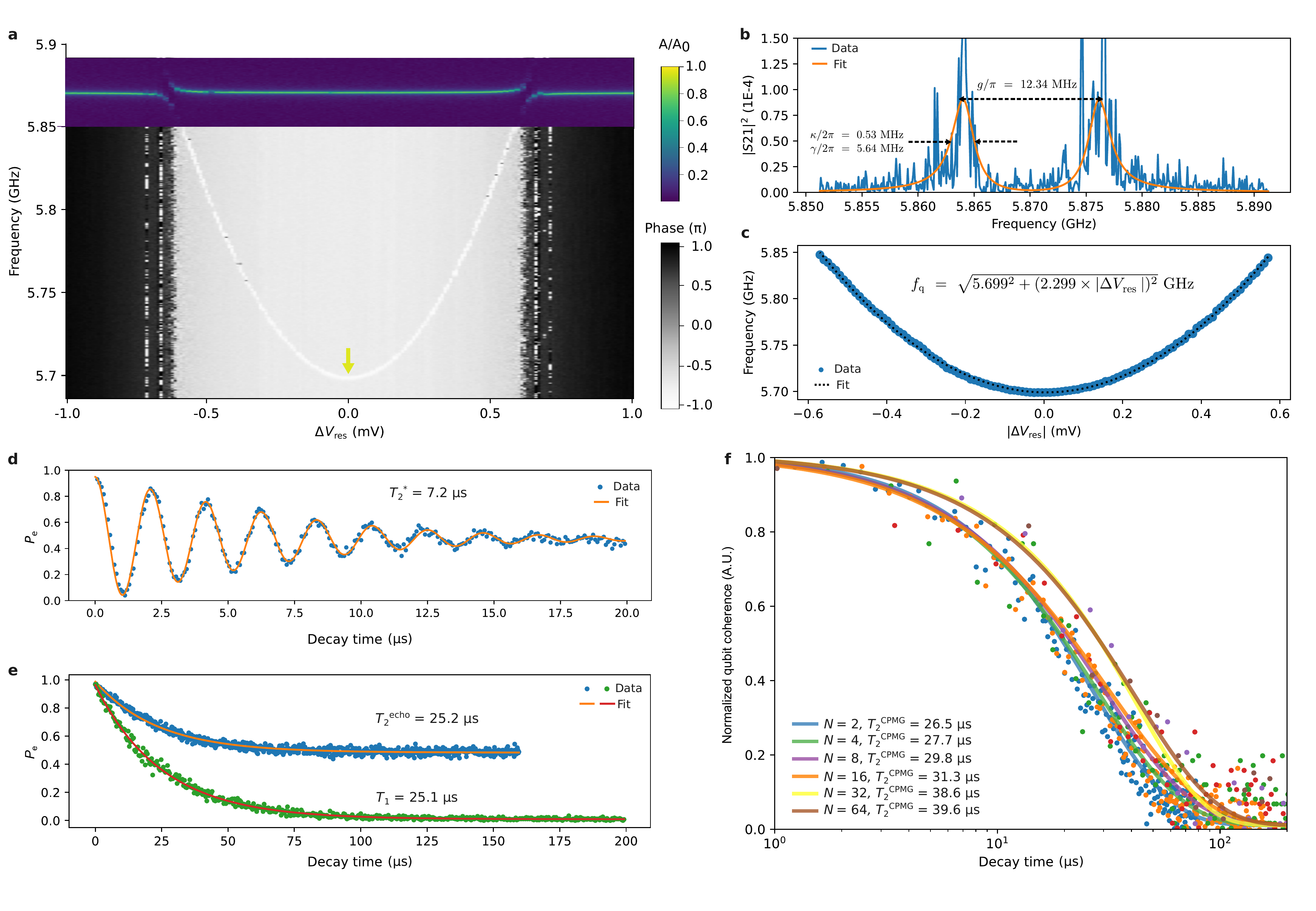}
	\caption{\textbf{Qubit 2 spectroscopy and coherence at charge sweet spot} \textbf{a}, Vacuum Rabi splitting and qubit spectrum as the function of relative resonator bias voltage $\Delta V_\text{res}$. \textbf{b}, Measurement and fitting of transmission spectrum ($\mathrm{|S21|}$) when Q2 is in resonance with the resonator. $g$, $\kappa$ and $\gamma$ are the qubit-resonator coupling strength, resonator and qubit linewidth, respectively. \textbf{c}, Fitting of the qubit spectrum ($f_\mathrm{q}$) to the model in the main text. \textbf{d}, Ramsey fringes at the charge sweet spot to extract the total dephasing (Ramsey) time $T_2^*$. $P_\mathrm{e}$ represents the qubit excited state population. \textbf{e}, Energy relaxation and Hahn echo measurement at the charge sweet spot. $T_1$ and $T_2^\mathrm{echo}$ are the relaxation time and total decoherence time with one Hahn echo, respectively. \textbf{f}, Coherence time extended by dynamical decoupling with CPMG pulse sequence at the charge sweet spot. $N$ is the number of refocusing pulses in the dynamical decoupling sequence. $T_2^\mathrm{CPMG}$ is the extracted total decoherence time under CPMG sequences.}
	\label{Fig:Q2_spec}
\end{figure}
\subsection{Spectroscopic and coherence characterization}

Figure S5a plots the vacuum Rabi splitting (top panel) and the two-tone qubit spectrum versus the relative bias voltage on the resonator $\Delta V_\text{res}$. Same as Q1, Q2 is insensitive to voltage bias applied on the resonator and trap guard, as shown in Fig. 1 of the main text. By fitting the on-resonance transmission spectrum, as shown in Figure S5b, we extract the qubit-resonator coupling strength of $g/2\pi~=~6.17$\,MHz and qubit linewidth of $\gamma/2\pi~=~5.84$\,MHz. Figure S5c plots the extracted qubit spectrum of $f_\text{q} ~=~ \sqrt{(5.699\times10^9)^2+(1.516\times10^9\Delta V_\text{res})^2}$, with $f_\text{q}$ in unit of Hz and $\Delta V_\text{res}$ in unit of mV. Figure S5d-e show the measured $T_\text{1}~=~25.1$\,\textmu s, $T_\text{2}^{*}~=~7.2$\,\textmu s, and $T_\text{2}^\text{echo}~=~25.2$\,\textmu s at the charge sweet spot. Figure S5f shows that with refocusing pulses, the $T_\text{2}^\text{CPMG}$ increases to approach $2T_\text{1}$ when biased at the sweet spot.

\begin{figure}[htb]
	\centering
	\includegraphics[scale=0.335]{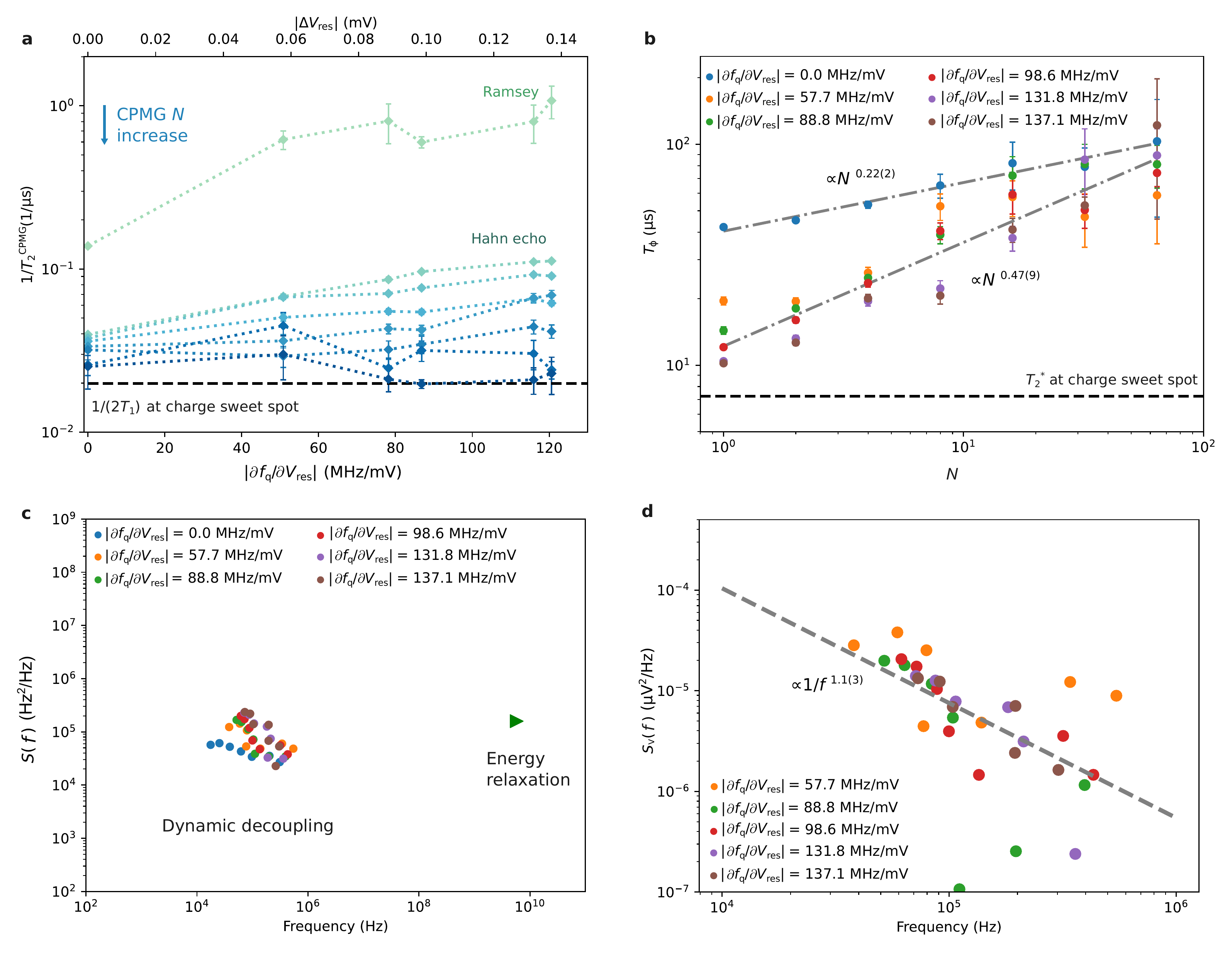}
	\caption{\textbf{Qubit 2 noise characterization with dynamical decoupling methods.} \textbf{a}, Extracted total decoherence rate under CPMG sequences $1/T_\text{2}^\text{CPMG}$ as the function of refocusing pulse number $N$ and qubit's sensitivity to voltage fluctuation $|\partial f_\text{q}/\partial V_\text{res}|$. $T_1$ is the qubit relaxation time. \textbf{b}, Fitted pure dephasing time $T_\mathrm{\phi}$ as the function of $N$ and $|\partial f_\text{q}/\partial V_\text{res}|$. \textbf{c}, Calculated total noise spectral density ($S(f)$, $f$ as frequency) via dynamical decoupling methods and transverse noise due to energy relaxation. \textbf{d}, Calculated equivalent voltage noise density ($S_\mathrm{v}(f)$ on the resonator electrode. All error bars represent the one standard error of extracted parameters (Methods).}
	\label{Fig:Q2_noise}
\end{figure}

\subsection{Dephasing time and noise spectroscopy}

Figure S6a plots the calculated $T_\text{2}^\text{CPMG}$ as a function of refocusing pulse number $N = 0,1,2,4,8,16,32,64$ for Q2 biased at different $\Delta V_\text{res}$ and sensitivity to voltage fluctuation $|\partial f_\text{q} / \partial \Delta V_\text{res}|$. With an increase in the refocusing pulse number, $T_\text{2}^\text{CPMG}$ all approach $2T_\text{1}$. When the qubit is biased to be more sensitive to charge noise, more refocusing pulses are needed to mitigate high-frequency noise to push $T_\text{2}^\text{CPMG}$ towards the relaxation limit. Figure S6b plots the fitted pure dephasing time $T_{\phi}$ as the function of $N$ and $|\partial f_\text{q} / \partial \Delta V_\text{res}|$. Unlike Q1 in the main text, Q2 suffers from more significant high-frequency noise when biased at its charge sweet spot, requiring more refocusing pulses to increase $T_{\phi}$. Besides, the power-law fittings for data acquired on and off the charge sweet spot are distinct. On the sweet spot, $T_{\phi}~\propto~N^{0.22(2)}$. In contrast, when away from the sweet spot $T_{\phi}~\propto~N^{0.47(9)}$. This observation indicates that the charge noise dominates the qubit decoherence when it is biased away from the charge sweet spot. As shown in Fig.\,S6c, we calculate the total noise measured by the dynamical decoupling method between 10\,kHz and 1\,MHz. Given the qubit's sensitivity $|\partial f_\text{q}/\partial V_\text{res}|$, and extracting the noise source contributing to the dephasing when biased at the sweet spot, we calculate the resonator electrode's equivalent voltage (charge) noise considering the qubit's sensitivity to voltage fluctuation. As shown in Fig.\,S6d, the calculated charge noise also falls in the range of $10^{-5}$\,\textmu $\text{V}^2/\text{Hz}$, with a fitted power-law of $S_v\propto f^{-1.1(3)}$.
 
\subsection{Temperature-dependent coherence}

Figure S7a plots the $T_\text{1}$ versus mixing chamber (MXC) temperature when Q2 is biased at its charge sweet spot. The blue dash curve shows the thermal decay model~\cite{ lisenfeld2010measuring, leggett1987dynamics}, with measured low-temperature $T_\text{1}$ at 10\,mK as $T_\text{1}(T=0)$ and projection to the high-temperature values. At 300\,mK, Q2 maintained a $T_\text{1}~=~9.31$\,\textmu s, about 40\% of the low-temperature value. Meanwhile, the good match between the model and measured data, similar to Q1 in the main text, further confirms that thermal-induced decay dominates the degradation of $T_\text{1}$ at higher temperatures. Figure S7b plots the temperature-dependent $T_\text{2}^{*}$ and $T_\text{2}^\text{echo}$ of Q2 biased at its sweet spot. The power-law fittings give $T_\text{2}^{*}~\propto~1/T^{1.2(2)}$ and $T_\text{2}^\text{echo}~\propto~1/T^{1.7(1)}$. The difference in the fitted exponent indicates a faster increase in high-frequency noise components with higher temperatures than low-frequency components. Figure S7c plots the extracted pure dephasing time and the calculated resonator induced dephasing based on the same model in the main text~\cite{anferov2024superconducting, reagor2016quantum}. The dispersive shift of the resonator when Q2 is biased at its sweet spot is estimated by $\chi\simeq g^2/\Delta$, where $g$ is the qubit-resonator coupling strength, and $\Delta$ is the qubit-resonator detune~\cite{schuster2007circuit}. The model matches well to the high-temperature values of $T_{\phi}$.

\begin{figure}[htb]
	\centering
	\includegraphics[scale=0.2]{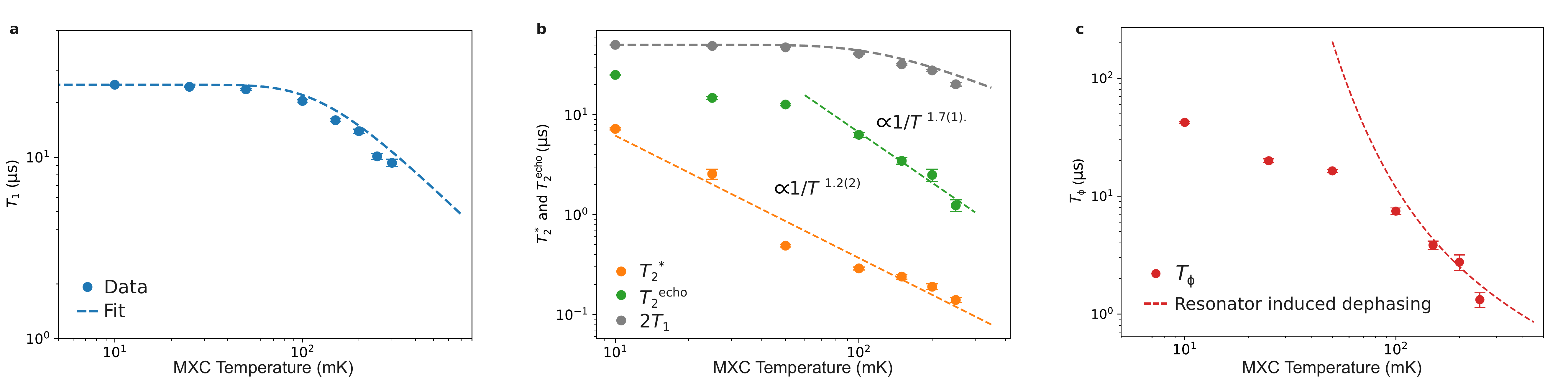}
	\caption{\textbf{Temperature-dependent coherence of Qubit 2.} \textbf{a}, Qubit relaxation time $T_\text{1}$ versus mixing chamber (MXC) temperature, with projection based on the thermal decay model $T_\text{1}(T) = T_\text{1}(T=0)\cdot\text{tanh} (\hbar \omega_{\rm{q}}/2k_\text{B}T)$. We use the measured $T_\text{1}$ at 10 mK as $T_\text{1}(T=0)$. $T$ is the MXC temperature. \textbf{b}, Total dephasing (Ramsey) time $T_\text{2}^{*}$ and total decoherence time under one Hahn echo $T_\text{2}^\text{echo}$ versus MXC temperature, with power-law fitting. Gray dots and the dashed curve show the value of $2T_\text{1}$. \textbf{c}, Pure dephasing time $T_{\phi}$ versus MXC temperature, with a parameter-free calculation based on resonator induced dephasing model. All error bars represent the one standard error of extracted parameters (Methods).}
	\label{Fig:Q2_temp}
\end{figure}

\newpage

\section{Qubit 3 (Q3) supporting data}

Q3 was trapped on the same resonator used to trap Q1 in the main text. During its short trapping lifetime, we managed to characterize its coherence properties at the charge sweet spot. This section will focus on reporting the relevant data.

\subsection{Spectroscopic and coherence characterization}

Figure S8a plots the vacuum Rabi splitting versus the relative bias voltage on the resonator $\Delta V_\text{res}$. Figure S8b shows the zoom-in plot of right splitting when adjusting the resonator guard electrode. The fit of avoided crossing gives a coupling strength of $g/2\pi~\simeq~16$\,MHz. We could not resolve the two transmission peaks when the qubit is in resonance with the resonator, indicating a relatively large on-resonance qubit linewidth. Figure S8c plots the theoretical calculation of the transmission spectrum with a fitted $g$ and a qubit linewidth of $\gamma/2\pi = 15$\,MHz, which qualitatively matches the experimental measurements. Figure S8d plots the two-tone measured qubit spectrum with a charge sweet spot at 4.456\,GHz. In Figure S8e-g, we show the measured $T_\text{1}~=~74.3$\,\textmu s, $T_\text{2}^{*}~=~3.46$\,\textmu s, and $T_\text{2}^\text{echo}~=~3.96$\,\textmu s at the charge sweet spot. With the large qubit-resonator detuning, the energy relaxation is dominated by non-radiative decay at the sweet spot.

\begin{figure}[htb]
	\centering
	\includegraphics[scale=0.35]{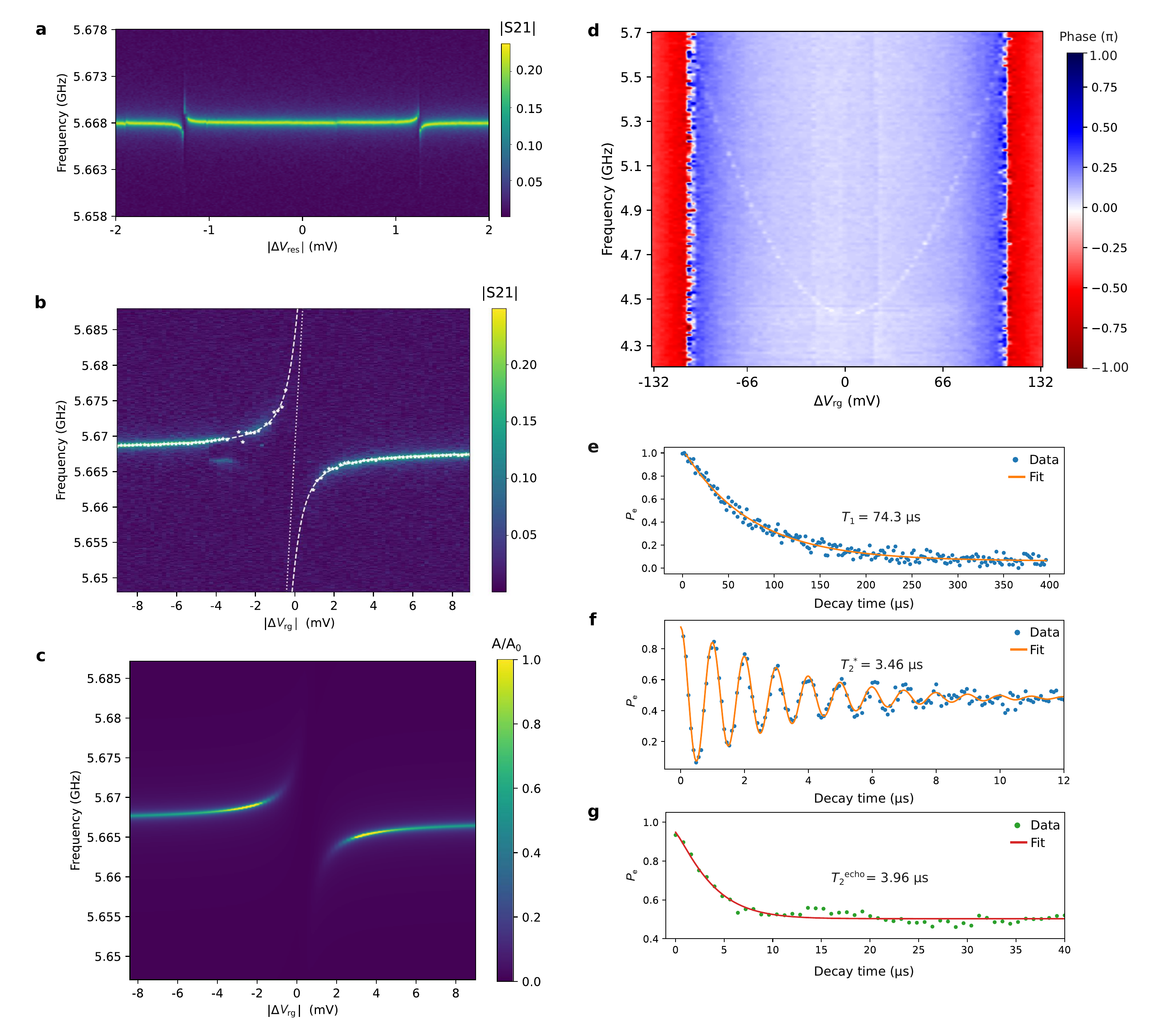}
	\caption{\textbf{Qubit 3 spectroscopy and coherence at sweet spot.} \textbf{a}, Vacuum Rabi splitting. $\mathrm{S21}$ is the measured transmission amplitude. $\Delta V_\mathrm{res}$ is the relative bias voltage on the resonator electrode. \textbf{b}, Fitting of qubit-resonator coupling strength $g$. White dots and dashed curves show extracted transmission peaks and fit to the two branches of the splitting. The white dot-dashed line shows the linear fitting of the qubit spectrum near the resonator frequency. $\Delta V_\mathrm{rg}$ is the relative bias voltage on the resonator gate electrode. \textbf{c}, Calculation of resonator transmission spectrum with parameters acquired from fitting in b and qubit linewidth $\gamma/2\pi~=~15$\,MHz. \textbf{d}, Two-tone measured qubit spectrum as the phase response at the resonator frequency. \textbf{e}, Energy relaxation time $T_1$, \textbf{f}, Total dephasing (Ramsey) time $T_2^*$, and \textbf{g}, Total decoherence time with one Hahn echo $T_2^\mathrm{echo}$ at the charge sweet spot. $P_\mathrm{e}$ is the qubit excited state population.}
	\label{Fig:Q3_spec}
\end{figure}

\newpage

\subsection{Dephasing time and noise spectroscopy}

We managed to improve the coherence time of Q3 at its sweet spot with the dynamical decoupling method. As shown in Fig.\,S9a-b, $T_\text{2}^\text{CPMG}$ increased to be above 100\,\textmu s when 64 refocusing pulses are applied. Similar to Q2, Q3 experienced significant high-frequency noise components when biased at its sweet spot. The inconsistency in total noise distribution on the charge sweet spot among the three qubits indicates the complexity of noise sources for the eNe charge qubit. Figure S9c plots the fitted $T_{\phi}$ as the function of $N$, with a power-law fitting of $T_{\phi}~\propto~N^{0.77}$. Figure S9d shows the extracted total noise spectral density near the 0.1\,MHz range, with similar absolute values as Q1 and Q2. 

\begin{figure}[htb]
	\centering
	\includegraphics[scale=0.35]{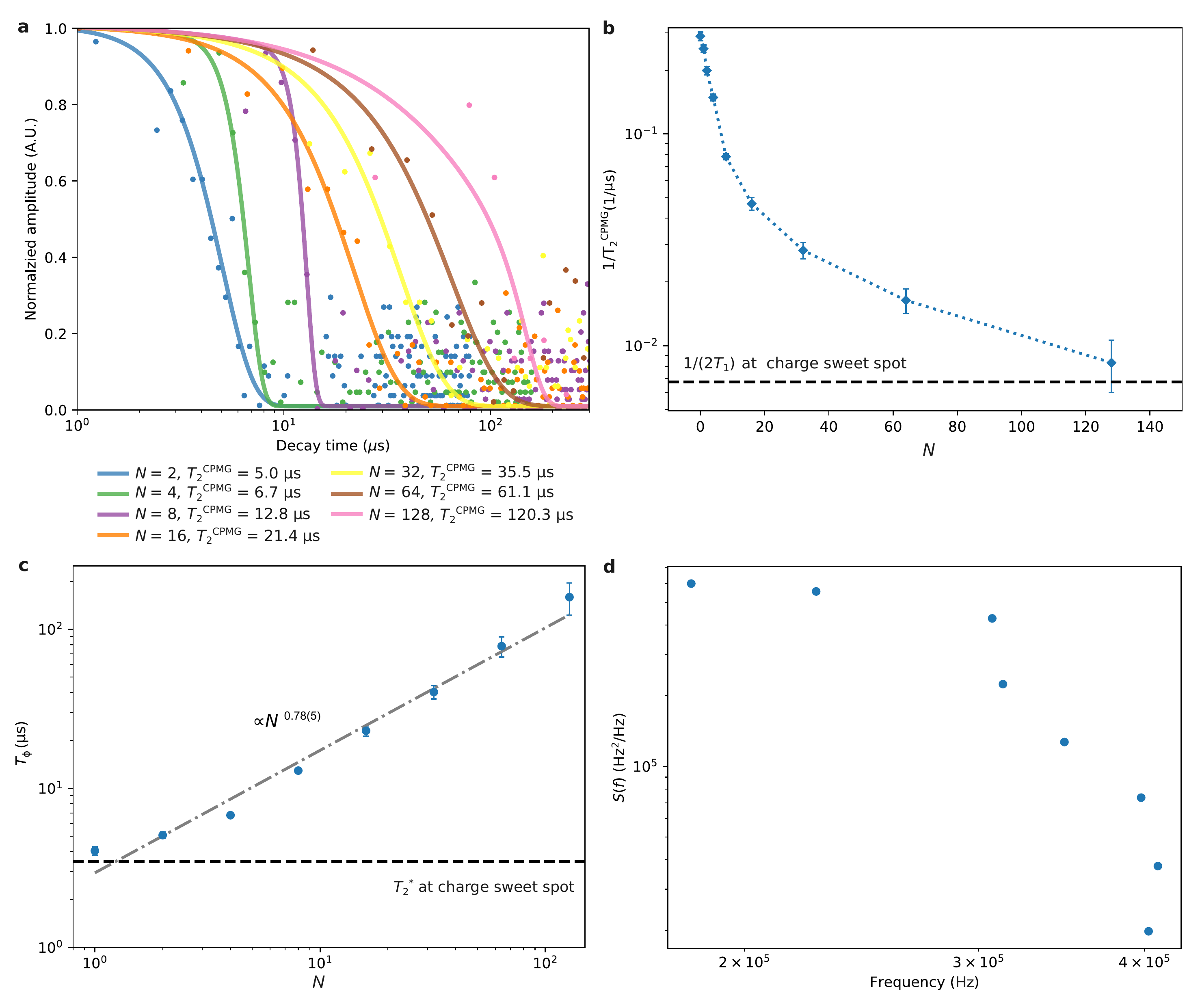}
	\caption{\textbf{Dynamical decoupling and noise characterization of Qubit 3 at the sweet spot.} \textbf{a}, Extension of coherence time via dynamical decoupling with CPMG pulse sequence. $N$ is the number of refocusing pulses in the dynamical decoupling sequence. $T_2^\mathrm{CPMG}$ is the extracted total decoherence time under CPMG sequences. \textbf{b}, Calculated $1/T_\text{2}^\text{CPMG}$ as the function of refocusing pulse number $N$. \textbf{c}, Fitted pure dephasing time $T_{\phi}$ as the function of $N$. $T_2^*$ is the total dephasing (Ramsey) time. \textbf{d}, Calculated total noise spectral density $S(f)$ near frequency ($f$) of 0.1\,MHz, characterized via dynamical decoupling methods.}
	\label{Fig:Setup}
\end{figure}

\newpage

\section{Spin coherence estimation}

In this section, we estimate the spin echo coherence of eNe spin qubits. Spin-nuclear interaction and the conversion of charge noise to spin are the two major decoherence sources for electron spin qubits. In eNe spin qubits, the contribution of charge noise to spin decoherence depends on the specific implementation of the spin states. Ref.~\cite{tian2025nbtin} proposes to generate spin-charge conversion in eNe trapped in a hundred-nanometer-wide double-well potential with micro-magnets. The estimated high-frequency-noise-induced total spin linewidth $\gamma'_\text{s}$, contributed by the charge linewidth $\gamma_\text{c}$ at the charge sweet spot, is scaled by the mixing factor $\Lambda$, following
\begin{equation}
 \gamma'_\text{s}=\Lambda^2\gamma_\text{c}+(1-\Lambda^2)\gamma_\text{s}
\end{equation} where $\gamma_s$ is the pure spin linewidth. $\Lambda$ is determined by the Zeeman splitting frequency $b_{||}$, the local field gradient $\partial B_\text{z}/\partial \text{y}$, and the charge state frequency at the sweet spot (i.e. effective tunneling coupling $2t_\text{c}$).
Applying the parameters in Ref.~\cite{tian2025nbtin}, including $b_{||}=4.8$\,GHz, $\partial B_\text{z}/\partial \text{y} = 0.36$\,mT/nm, and $2t_\text{c} = 8.0$\,GHz, we obtain $\Lambda = 1.9$. On natural solid neon, the estimated pure spin echo coherence is 30\,ms, limited by spin-nuclear coupling~\cite{chen2022electron}. Given the measured typical eNe charge qubit echo coherence of $5\sim20$\,\textmu s, we arrive at the estimation of eNe spin echo coherence of $0.1\sim0.5$\,ms.
It is comparable to the state-of-the-art records of semiconductor electron spin qubits, as listed in the Tab. SII. The primary limiting factor is the transfer of the suboptimal charge decoherence to the spin degree of freedom through the synthetic spin-charge conversion~\cite{tian2025nbtin}.
\\

\begin{table*}[t]
	\centerline{
		\begin{tabular}{|l|l|l|l|l|l|l|}
			\hline
			Qubit platform & Host interface & Qubit frequency (GHz) & Spin echo decoherence time $T_2^\mathrm{echo}$ (ms) & Reference  \\ \hline
            
			eNe&Vacuum/Neon  & 4.8 (Estimated*)  &  $0.1\sim0.5$ (Estimated*) & This work \\  \hline
            
			Silicon MOS&Si/SiO$_2$ & $\sim2\times10^{-3}$  & $5\times10^{-3}$ & ref.~\cite{jock2022silicon} \\ 
            
			&Si/SiO$_2$ & 16.2 & $3.6\times10^{-2}$ & ref.~\cite{klemt2023electrical} \\ 

            &$^{28}$Si/$^{28}$SiO$_2$ & 17.1 & $0.2$ &ref.~\cite{zwerver2022qubits} \\ \hline
            
			Group IV&Si/SiGe & $5\times10^{-2}$  & $5\times10^{-4}$ &ref.~\cite{connors2022charge}  \\ 
            
            heterostructures&Si/SiGe & $18.98$ & $7\times10^{-2}$ &ref.~\cite{kawakami2016gate} \\

            &$^{28}$Si/SiGe  & $\sim5\times10^{-3}$ & $0.7$ & ref.~\cite{eng2015isotopically}  \\
            
			&$^{28}$Si/SiGe & $18$  & $0.1$ & ref.~\cite{yoneda2018quantum}  \\  
            
			&$^{28}$Si/SiGe  & $19.9$  & $0.1$ &ref.~\cite{struck2020low}  \\ 
             \cline{2-5}

            &Ge/SiGe  & $3.52$ & $3.8\times10^{-3}$ & ref.~\cite{hendrickx2021four} \\

            &Ge/SiGe  & $0.95$ & $3.6\times10^{-2}$ & ref.~\cite{stehouwer2024exploiting}  \\ 

            &Ge/SiGe  & $\sim3\times10^{-2}$ & $0.2$ & ref.~\cite{hendrickx2024sweet}  \\ \hline
            
            Group III-V&GaAs/AlGaAs & $6.1\times10^{-2}$ & $2\times10^{-4}$ & ref.~\cite{cerfontaine2016feedback}  \\

			heterostructures&GaAs/AlGaAs & $7\times10^{-2}$ & $9\times10^{-3}$ & ref.~\cite{dial2013charge} \\ 
            &GaAs/AlGaAs & 0.12 & $6.1\times10^{-3}$ & ref.~\cite{cerfontaine2020closed}  \\ \hline

            \hline

	  \end{tabular}
            }
	\caption{Summary of spin echo coherence obtained from various qubit platforms. *Estimated spin echo decoherence time of an eNe spin qubit in a double quantum dot configuration with micro-magnets, based on the charge noise characterized in this study.}
	\label{tab:Charge_noise_table}
\end{table*}

\section{System characterization before and after electron emission}

We run control experiments to show the effects of neon on electron trapping. Figure S10 plots the representative resonator spectroscopies while sweeping the resonator voltage for four situations, i.e., with or without neon deposition and before or after emitting electrons from the tungsten filament. Before firing the filament or without neon, we rarely observe strong couplings. We fire the filament following the same procedure as in the Methods of the main text. We observed a significant increase in both the number and strength of anti-crossing features over the same voltage range only when the resonator is covered with a thin layer of neon. We interpret these features as originating from both trapped eNe qubits and floating electrons with high mobility near the resonator electrodes. These results indicate that the neon layer enables the stable trapping of electrons strongly coupled with the resonator. For a typical experiment, we then zoom in to conduct refined scanning near the anti-crossings to characterize the qubit spectroscopy, followed by time-domain coherence measurements. Weak coupling features appeared after firing the filament without coating neon. One possible origin of that could be the penetration of high-energy electrons into the substrate Only when both the neon film is present on the sample surface and electrons are emitted, are we able to characterize eNe qubits with vacuum Rabi splitting, two-tone qubit spectroscopy, and time domain measurements. Future electron loading mechanisms, such as photoemission, could be gentler to reduce the weak coupling backgrounds.

\begin{figure}[]
	\centering
	\includegraphics[scale = 0.6]{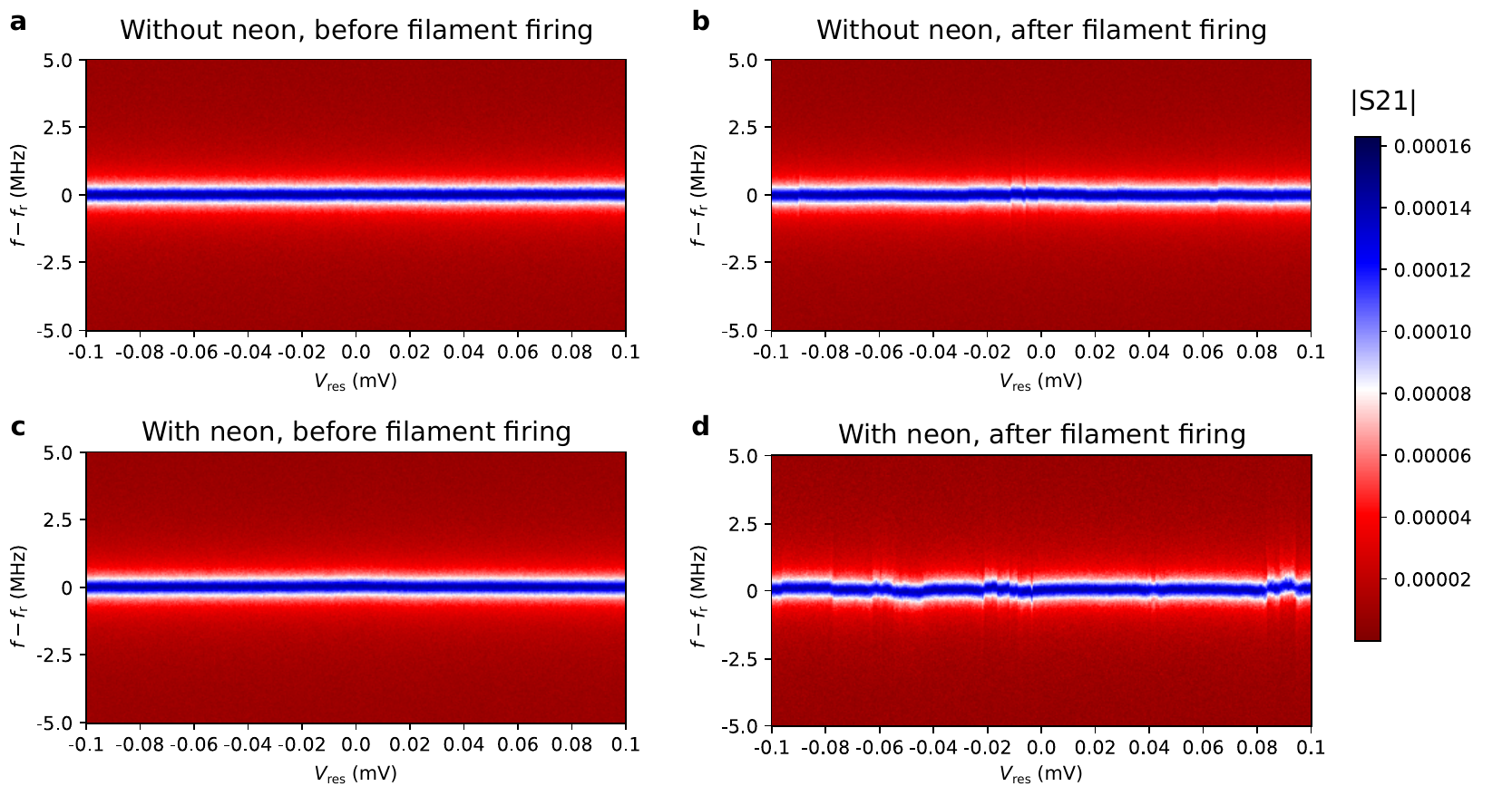}
	\caption{\textbf{Resonator response to electron emission.} Representative instance of resonator transmission spectroscopy versus bias voltage on resonator electrode $V_\text{res}$ without (\textbf{a}, \textbf{b}) and with (\textbf{c}, \textbf{d}) neon deposition, and before (\textbf{a}, \textbf{c}) and after (\textbf{b}, \textbf{d}) emitting electrons onto the device. Qualitatively, the number and strength of anti-crossing features increase under the condition after firing the filament on the neon-covered resonator. The y-axis denotes the frequency ($f$) difference from the resonator frequency $f_\mathrm{r}$. $\mathrm{|S21|}$ is the transmission amplitude.}
	\label{Fig:Emission}
\end{figure}

\bibliography{eNe_noise}